\documentclass[sigconf, nonacm]{acmart}
\usepackage{booktabs} 
\usepackage{tabularx} 
\usepackage{multirow}
\usepackage{enumitem} 
\usepackage{subcaption}
\usepackage{soul} 

\usepackage[utf8]{inputenc}
\usepackage{xcolor}
\usepackage{colortbl}

\definecolor{lightyellow}{RGB}{247, 209, 178}
\definecolor{lightblue}{RGB}{197, 225, 251}
\definecolor{sagegreen}{RGB}{138, 154, 91}
\definecolor{navyblue}{RGB}{34, 51, 84}
\definecolor{darkblue}{RGB}{82, 127, 137}
\newcommand{\HS}[1]{
  \colorbox{lightyellow}{\textcolor{black}{HS-P#1}}
  }
  
\newcommand{\MS}[1]{
  \colorbox{lightblue}{\textcolor{black}{MS-P#1}}
  }

  \newcommand{\ES}[1]{
  \colorbox{sagegreen}{\textcolor{white}{ES}}
  }

\newcommand{\HSD}[1]{
  \colorbox{lightyellow}{\textcolor{black}{HS}}
  }
  
\newcommand{\MSD}[1]{
  \colorbox{lightblue}{\textcolor{black}{MS}}
  }

\AtBeginDocument{%
  }

\begin{document}
\settopmatter{printacmref=false} 
\setcopyright{none}             
\renewcommand\footnote[1]{}     

\title{How Children Design and Reason about Trustworthy AI Chatbots}


\author{Deniz Ozturk}
\orcid{0009-0003-7245-7501}
\affiliation{%
  \institution{North Carolina State University}
  \city{Raleigh}
  \country{United States}}
\email{dozturk@ncsu.edu}

\author{Jiayu Li}
\affiliation{%
  \institution{Independent Researcher}
  \city{Raleigh}
  \country{United States}}
\email{jiayuli.tyler@gmail.com}

\author{Daksh Pratap Singh}
\orcid{0009-0001-4990-3345}
\affiliation{%
  \institution{North Carolina State University}
  \city{Raleigh}
  \country{United States}}
\email{dsingh23@ncsu.edu}

\author{Yasitha Rajapaksha}
\orcid{0000-0002-3346-6304}
\affiliation{%
  \institution{North Carolina State University}
  \city{Raleigh}
  \country{United States}}
\email{yrajapa@ncsu.edu}

\author{Fasika Melese}
\orcid{0000-0003-2737-2544}
\affiliation{%
  \institution{University of Pennsylvania}
  \city{Philadelphia}
  \country{United States}}
\email{fasikaye@upenn.edu}

\author{Bahare Riahi}
\orcid{0009-0005-4560-4857}
\affiliation{%
  \institution{North Carolina State University}
  \city{Raleigh}
  \country{United States}}
\email{briahi@ncsu.edu}

\author{Shiyan Jiang}
\orcid{0000-0003-4781-846X}
\affiliation{%
  \institution{University of Pennsylvania}
  \city{Philadelphia}
  \country{United States}}
\email{jiang33@upenn.edu}

\author{Qiao Jin}
\orcid{0000-0001-5493-1343}
\affiliation{%
  \institution{North Carolina State University}
  \city{Raleigh}
  \country{United States}}
\email{qjin4@ncsu.edu}

\author{Joey Huang}
\orcid{0000-0003-4362-326X}
\affiliation{%
  \institution{North Carolina State University}
  \city{Raleigh}
  \country{United States}}
\email{joeyhuang@ncsu.edu}

\author{Veronica Cateté}
\orcid{0000-0002-7620-7708}
\affiliation{%
  \institution{North Carolina State University}
  \city{Raleigh}
  \country{United States}}
\email{vmcatete@ncsu.edu}

\author{Tiffany Barnes}
\orcid{0000-0002-6500-9976}
\affiliation{%
  \institution{North Carolina State University}
  \city{Raleigh}
  \country{United States}}
\email{tmbarnes@ncsu.edu}

\author{Xiaoyi Tian}
\orcid{0000-0002-5045-0136}
\affiliation{%
  \institution{Kennesaw State University}
  \city{Marietta}
  \country{United States}}
\email{xtian5@kennesaw.edu}

\renewcommand{\shortauthors}{Ozturk et al.}





\begin{abstract}
    Children increasingly interact with AI chatbots, making trust calibration essential to AI literacy. Prior research has examined children's trust in AI mainly as \textit{users} evaluating systems built by others, rather than as designers of their own chatbots. We developed a chatbot-building environment with adjustable trust-relevant traits (e.g., confidence, transparency, formality, assertiveness), rules, and persona. We conducted mixed-methods study with 115 learners (ages 8--18) who made 119 chatbots. We examined how children configured their chatbots, reasoned about trustworthiness, and how closely chatbot behavior aligned with their designs. Younger students (age 10-13) set significantly higher confidence than older students (age 14-18), and some deliberately built chatbots that gave wrong answers on purpose, yet still called them trustworthy, arguing that a chatbot does what it was built to do. Younger students equated trust with purpose-fulfillment, while older students linked it to transparent, calibrated design. Students also calibrated academic chatbots to be more transparent and formal than hobby chatbots. We identify seven design dimensions describing what children believe makes a chatbot trustworthy, and discuss implications for AI literacy tools.
    
\end{abstract}

\maketitle

\section{Introduction}

Children increasingly interact with AI chatbots in their everyday lives, using them for homework help, entertainment, and information seeking \cite{kuhail2022chatbots, riahi2026chatbots}. Because chatbots can produce inaccurate or misleading responses, children must decide when their information is trustworthy, often without sufficient knowledge or strategies to evaluate it \cite{tian2026hallucination}. Overtrust can leave children with incorrect information they may not know to question, whereas undertrust can prevent them from benefiting from AI tools that could support learning and exploration \cite{oh2025hey, solyst2024childrenovertrust, movahed2025ask}. Supporting appropriately \textit{calibrated trust} therefore requires understanding what children believe makes AI trustworthy \cite{dubiel2022trustcalibration, riahi2026humanizing}.

Most research on children's trust in AI, positions children as \textit{users} who evaluate systems designed by others \cite{vanbrummelen2023children, oh2025hey}. This perspective helps explain how children judge existing AI systems, but tells us less about how children think trustworthiness should be designed into those systems. Constructionist research extends this perspective by allowing children to program conversational agents \cite{tian2023amby, lin2020zhorai, van2021alexa} and examining changes in their perceptions of systems such as Alexa \cite{van2021alexa, vanbrummelen2024learning}. Related work also investigates children's ideas about ideal conversational agents \cite{vanbrummelen2023children}. However, we still know less about how children translate their ideas about trustworthiness into chatbots they personally create. What do children believe a trustworthy chatbot should do, and how do they attempt to build those qualities into its behavior?

To understand that perspective, we position children as \textit{designers} who make decisions about how an AI system should behave. When children take on the role of designers, their ideas about trustworthiness can become more visible as the design process requires children to translate those ideas into concrete choices. Examining those choices can reveal what children already understand about trustworthy AI, as well as which aspects of trustworthiness may still require further reflection, guidance, or support \cite{druga2021children, vanbrummelen2024learning}.
Chatbot design provides a particularly useful setting for studying this process because conversational AI is becoming an increasingly accessible form of AI for children to create and customize \cite{tian2023amby, brummelen2020teaching}.

Prior work on chatbot design has examined how adjustable chatbot characteristics, feedback, and different use cases can shape children's trust in conversational systems \cite{vereschak2024trust, oh2025hey, vanbrummelen2024learning}. However, these studies tell us less about how children themselves make trust-related design decisions when building chatbots.

Our study is guided by the following research questions:

\begin{itemize}
    \item[\textbf{RQ1.}] What kinds of chatbots did students build, and how did they configure them?
    \item[\textbf{RQ2.}] What design choices do students believe make an AI chatbot trustworthy?
    \item[\textbf{RQ3.}] How closely did student-created chatbots align with their intended design?
\end{itemize}

To investigate our research questions, we conducted a mixed-methods study with 115 students--76 students at Grades 2-3 (ages 8-9), 24 students at Grades 6-7 (ages 10-13), 15 students at Grades 9-12 (ages 14-18)--, who used our chatbot-building environment in a workshop to design, build, and iteratively refine their own chatbots. 

We combine quantitative analysis of pre/post surveys and tool logs with qualitative thematic coding of interviews and design artifacts.

The primary contributions of our work is as follows: 

\begin{enumerate}[label=\textbf{(\arabic*)}, leftmargin=*, labelwidth=2cm, align=left, labelsep=0.5em] 

\item \textbf{New understanding about how children conceptualize trustworthiness when they design AI chatbots.} We show that children treat trust as the result of multiple interacting design choices. We identify seven design dimensions capturing how children think about AI trustworthiness when building their own chatbots.
This gives future researchers a direction for studying child AI trust and building age appropriate features for trustworthy AI design.

\item \textbf{Exploratory evidence that age groups and domain shape trust design across independent measures.} Across configuration logs, stated beliefs and design fidelity, we examine whether trust design varies by age groups and by chatbot domain (academic vs hobby-driven). Domain differences were consistent across all four measures. We also conducted exploratory analysis on different design and reasoning patterns among younger children and older children groups. 

\item \textbf{Design implications for constructionist AI literacy platforms.} We identify opportunities to help children articulate their intentions, test how their choices affect chatbot behavior, and revise their designs when a chatbot fails to demonstrate the qualities they intended. This informs age-appropriate support for children as designers of trustworthy AI.
  
\end{enumerate}

\section{Related Work}

\subsection{Children's Trust in AI and Conversational Agents}

Trust in AI is often defined as the belief that a system will help someone reach their goal despite some uncertainty \citep{lee2004trust}. Researchers distinguish between \textbf{trustworthiness}, a system's actual qualities, and \textbf{trust}, how people perceive them \citep{ragone2026children}. \citet{vanbrummelen2023children} applied an established human-trust framework and found that children most often referenced competence and predictability when evaluating conversational agents like Alexa or Google.  

Children often overtrust conversational agents, attributing human-like qualities to them and judging them as more knowledgeable and reliable than they are, especially when the systems are less transparent about their limits \citep{oh2025hey, movahed2025ask, andries2023alexa}.

Transparency about how an agent works can help calibrate this trust \cite{vossing2022designing, vereschak2024trust, kim2024imnotsure}. \citet{vanbrummelen2023children} found that children's trust shifted most when they learned about an agent's information sources. \citet{bailey2021perspective} extend this further, arguing that the datasets and design underlying children's conversational agents should be curated with children's ethical and cultural context in mind. \citet{li2026supporting} found that learners using generative chatbots often engaged shallowly and missed errors due to overtrust and cognitive overload, but when the chatbot expressed uncertainty, learners adopted incorrect answers less often. This suggests that design choices have measurable effects on how users calibrate their trust.

Most of this work, however, has two limitations. First, trust is typically measured through post-interaction surveys or binary judgments (e.g., thumbs up/down), which capture whether children trust an agent but how they reasoned about that judgment, and most existing scales were designed for adults, not children \citep{oh2025hey, ragone2026children}. Second, children are positioned as users evaluating a system built by others (e.g., Alexa), not as designers making trust-related decisions themselves \cite{vanbrummelen2023children, oh2025hey, andries2023alexa, movahed2025ask}. Our study addresses this gap directly by examining the design choices and explanations children give when they build the agent themselves, revealing what they believe makes a chatbot trustworthy and how they try to build that trustworthiness in.

\subsection{Child-AI Interaction}

Children's interactions with AI are fundamentally shaped by developmental processes that influence how they perceive, trust, and learn from AI systems. Even very young children prefer reliable computers over unreliable ones \cite{geng2025living}. Children's trust shifts with age. Younger children are easily influenced by design features that make machines seem human-like, leading them to treat AI as a social agent \cite{hoehl2024honest, nass1994computers, geng2025living}. \citet{hoehl2024honest} found that older children are less susceptible to this anthropomorphism, but they are actually \textit{more} likely to trust the information machines provide. Consistent with this, children aged 6--11 overestimate voice assistants' intelligence \cite{andries2023alexa}, while school-age children show greater skepticism of inaccurate robots than adults, increasing with age \cite{flanagan2024school}.

Prior work shows that actively engaging with how AI works can recalibrate children's trust. \citet{druga2021children} found that having children train custom machine learning models and integrate them into coding environments increased their skepticism of device capabilities and shifted their attribution of agency from the machine toward its human programmers \cite{druga2017hey}. Similarly, \citet{vanbrummelen2024learning} found that teaching youth how agents are programmed, where they get information, and how they process input led participants to articulate changes in trust around predictability, competence, and integrity. However, these interventions do not examine how children's own design choices vary across developmental stages, or how children reason about trust when the interface itself is theirs to configure.

\citet{fusco2026towards} raise this directly in the context of child-AI co-creation, identifying open questions about whether collaboration patterns and communication strategies differ by age.

\citet{cai2025child} synthesize this landscape into design principles for child-AI co-creation tools, including Appropriate Reliance, Peer/Family Collaboration, and Process Transparency. We implement these through Fact Check feature, Confidence Level, shareable Peer Testing, adjustable temperature and model selection, respectively.

\subsection{AI Literacy through Chatbot Development}

AI literacy has emerged as a priority for preparing young people to critically engage with artificial intelligence \citep{long2020ai, ng2021ai, druga2021children}. \citet{long2020ai} define it as a set of competencies for critically evaluating AI, communicating and collaborating with it, and using it as a tool; \citet{chiu2024artificial} distinguish AI literacy (knowledge) from competency (confident application). AI4K12's ``Five Big Ideas'' offers a curricular foundation, and work on how these ideas should be sequenced developmentally echoes our own emphasis on age as a design variable \citep{touretzky2019envisioning, druga2021children}. Across frameworks, technical understanding is consistently paired with ethics, critical reflection, and responsible use.

Chatbots are a particularly accessible entry point for this literacy, since children already encounter conversational agents in everyday contexts, from voice assistants to educational tools \citep{kuhail2022chatbots, riahi2026chatbots}. But these systems can generate plausible but incorrect information, or hallucinations \citep{huang2025hallucinations}, which can spread misinformation if left unchecked \citep{solyst2024childrenovertrust, danry2025misinformation}. Helping children understand why hallucinations occur and how to detect them is thus a core AI literacy skill.

Building a chatbot engages several of these literacies at once, from setting rules and constraints to reasoning about whether the chatbot can be trusted \citep{touretzky2019envisioning}. Chatbot design is thus a holistic entry point that connects multiple AI concepts through one hands-on activity, in the spirit of Constructionism \citep{papert1980}. Tools like Teachable Machine \citep{carney2020teachable}, Cognimates \citep{druga2018growing}, and Machine Learning for Kids \citep{lane2018mlforkids} have made this kind of hands-on engagement possible for learners.

Recent work has explored engaging youth in designing AI applications \citep{morales2025learning, tseng2024coml, tian2023amby}, but rarely asks how this shapes children's understanding of trust specifically. Our study fills this gap by positioning children as designers of their own chatbots and asking what they believe makes a chatbot trustworthy, and how they adjust concrete design features to act on those beliefs. In doing so, we extend AI literacy from a focus on using AI responsibly toward designing AI responsibly: understanding hallucinations is not only about detecting them as a user, but about knowing what design choices make them more or less likely to mislead someone else.

AI literacy and trust are also closely linked: teaching AI literacy in ways tailored to a specific audience helps calibrate how much that audience trusts a system \citep{vereschak2024trust}, and \citet{dangol2025beyond} treat trust directly as part of what AI literacy should teach children, positioning them not only as interpreters and collaborators of AI, but as \textit{resistors} who critically evaluate and reject outputs they judge untrustworthy. Designing a chatbot brings these together: it asks children to practice AI literacy by building something trustworthy, not only by evaluating outputs someone else built.

\section{Chatbot Development Environments}

\subsection{Design Goals and Interaction Workflow}
We designed \textit{LUMI} to position children as active designers of conversational AI (Figure \ref{fig:panel}). Drawing on Constructionist principles \cite{papert1980}, we reasoned that building a system, rather than just using it, would surface children's mental models of trustworthiness as design choices. Three goals guided our design: (1) making trust-relevant design choices explicit and adjustable, (2) enabling iterative refinement through testing, and (3) supporting peer feedback and sharing to foster reflection on trust.
The interaction workflow followed a structured \textit{configure-test-evaluate-revise} cycle. Students first configured their chatbot's identity, personality, rules, and communication traits. They then tested it by sending messages and inspecting responses. Next, they evaluated the output against their expectations and peer feedback. Finally, they revised their configurations and repeated the cycle. This cycle was designed to make trust reasoning visible at each stage: what children chose to configure, how they reacted to their chatbot's behavior, and what they changed in response.

\subsection{Key Components}
\label{sec:key_components}

\subsubsection{Configuring Chatbot Identity and Behavior}
Students defined their chatbot through a configuration panel that combined text fields, trait sliders, and a knowledge base upload area. The text fields captured: Bot Name, Description, Purpose (what the chatbot does and who it serves), Character \& Personality (the chatbot's persona), Communication Tone (how it speaks), and Rules \& Guidelines (topic boundaries, hint behavior, and handling of out-of-scope questions). These fields were designed to make students' design intentions explicit before they ever tested the system to understand their assumptions about what makes a trustworthy chatbot. 

Four trait sliders (Confidence, Transparency, Formality, and Assertiveness) controlled how the chatbot expressed itself. Each slider had Low, Medium, and High levels (1-3 Scale), which the system converted into written instructions for the underlying model. We designed these traits based on prior work on trust in automation \cite{lee2004trust, hoff2015trust}, selecting dimensions that prior research had shown to influence how users perceive system reliability and credibility.

\textbf{Confidence} controlled how definitively the chatbot phrased its answers. At the Low level, it hedged explicitly (e.g., "\textit{I'm not sure, but...}"). At the Medium level, it suggested answers while prompting the student to verify (e.g., "\textit{You may want to verify...}"). At the High level, it asserted answers directly (e.g., "\textit{The answer is...}" or "\textit{You should...}"). We included this trait because prior work has shown that perceived certainty strongly influences children's trust judgments \cite{li2026supporting}, and we wanted students to be able to manipulate this dimension directly.

\textbf{Transparency} controlled how openly the chatbot explained its own limitations and reasoning. At the Low level, it provided answers without context or qualification. At the Medium level, it acknowledged its limits only when directly relevant to the student's prompt. At the High level, it proactively clarified its scope, data sources, and any underlying uncertainty. We included this trait because transparency has been identified as a key factor in calibrating children's trust in AI systems \cite{vanbrummelen2023children, vossing2022designing}.

\textbf{Formality} controlled the tone of the chatbot's language, ranging from casual and conversational to formal and precise. \textbf{Assertiveness} controlled how strongly the chatbot stood behind an answer when the user disagreed, from deferring to the user to maintaining its position unless shown a contradictory source. Both traits were included to give students granular control over the chatbot's responses.

Students could optionally upload reference materials (e.g., documents, slides, spreadsheets, and images) to ground the chatbot's responses in provided sources. This knowledge base feature was designed to let students experiment with how verifiable sources affect perceived trustworthiness, a dimension that prior work has shown to be important in children's trust evaluations \cite{vanbrummelen2024learning}.

\subsubsection{Testing and Inspecting Responses}

The testing panel allowed students to send messages to their configured chatbot and inspect responses. Students could select from six language models (\cite{gpt54}, \cite{gpt54_mini}, \cite{claude_opus_46}, \cite{claude_sonnet_46}, \cite{claude_haiku_46}, \cite{llama32_3b_instruct})to observe how model choice affected output quality, and adjust a Temperature control that varied how predictable or expressive the model's wording was. A Tools menu enabled file search (using uploaded materials) and web search.

Each response displayed three parts: a Reasoning section showing the model's rationale, the Answer itself, and a Confidence bar (0--100) indicating the model's confidence in its answer. The Confidence bar was deliberately separate from the Confidence trait slider: the slider controlled how the chatbot sounded, while the bar reflected how confident the model was in its response. This distinction was designed to help students notice the gap between confident-sounding language and actual reliability.

\begin{figure*}[t]
  \centering
  \includegraphics[width=1\linewidth]{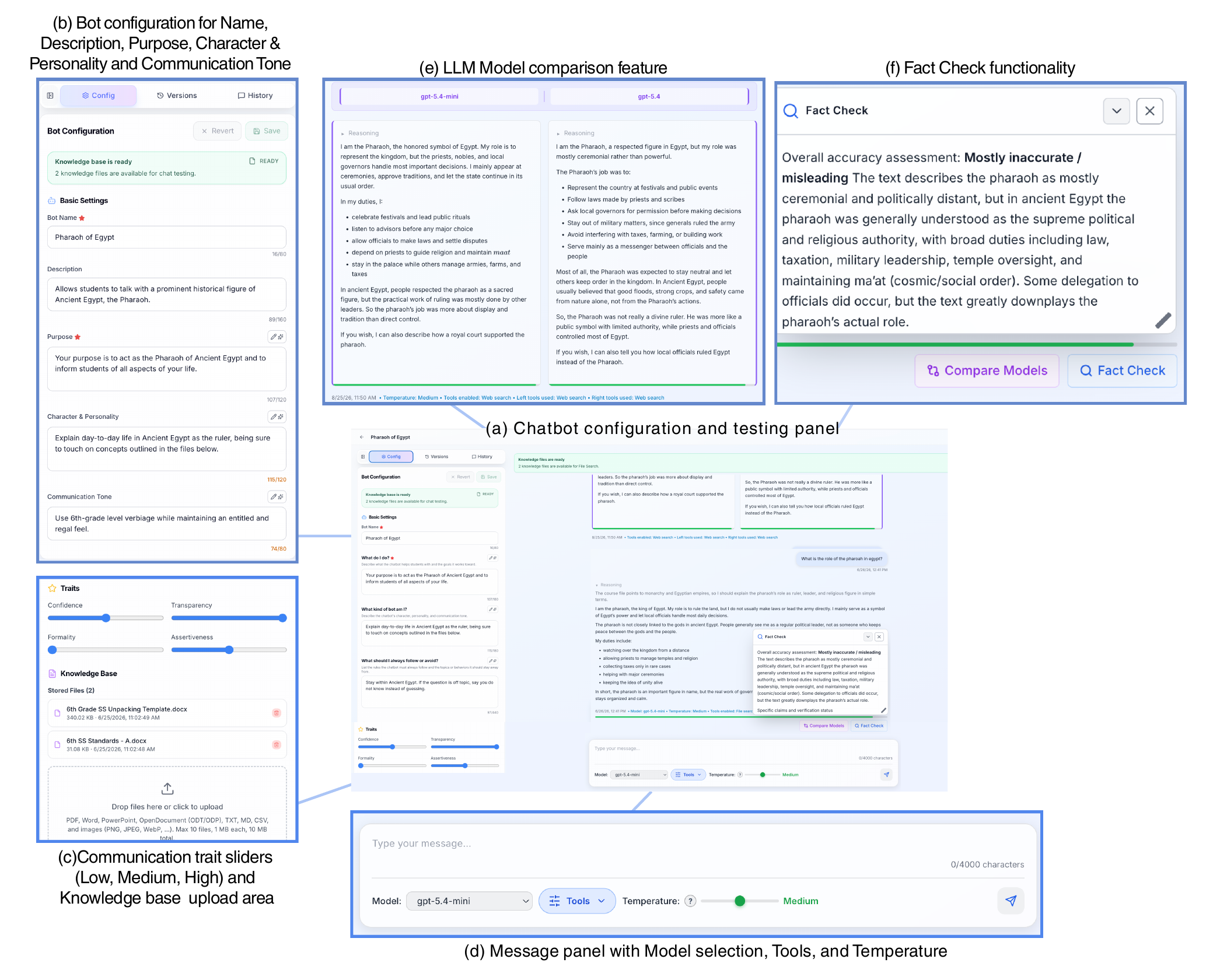}
  \caption{The LUMI interface. (a) The chatbot configuration panel. (b) Bot Configuration for basic settings (Name, Description, Purpose, Character \& Personality, Communication Tone and Rules \& Guidelines. (c) Trait sliders (Confidence, Transparency, Formality, Assertiveness) and Knowledge base, (d) Message panel with Model selection, Tools (file search and web search capabilities), and Temperature (low, medium, high).}
  \Description{(a) The chatbot configuration panel. (b) Bot Configuration for basic settings (Name, Description, Purpose, Character \& Personality, and Communication Tone). (c) Trait sliders (Confidence, Transparency, Formality, Assertiveness) and Knowledge base, (d) Message panel with Model selection, Tools (file search and web search capabilities), and Temperature (low, medium, high).}
  \label{fig:panel}
\end{figure*}

Two inspection features supported critical evaluation. \textbf{Fact Check} sent the reply to a separate model, which reviewed each claim and suggested corrections, providing a second opinion. \textbf{Compare Models} sent the same question to two models side-by-side, letting students judge which response was more accurate (see Figure~\ref{fig:panel} (e)). Both features were designed to scaffold students' ability to verify and critique AI outputs, addressing prior work showing that children often lack strategies to evaluate AI reliability \cite{tian2026hallucination, li2026supporting}.

\subsubsection{Revising and Sharing Chatbots}

Every save created a new, restorable version, letting students track their design progression and revert to any earlier configuration. Every test conversation remained viewable even after later changes, enabling us to capture each chatbot's full revision history as students iterated toward a version they considered trustworthy. Students could also share their chatbot with a peer to test, surfacing mismatches between design intent and user experience that often prompted revision. All configurations, conversations, and interaction events were timestamped for analysis.

\subsection{Controlled Hallucination}
\label{sec:hallucination}
\begin{figure*}[t]
  \centering
  \includegraphics[width=\linewidth]{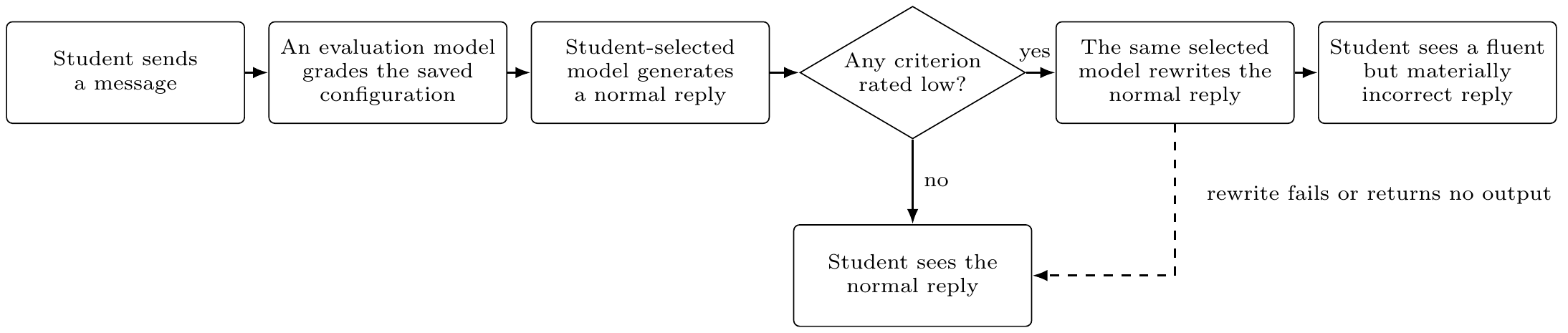}
  \caption{The controlled hallucination mechanism, showing the normal, rewriting, and fallback paths.}
  \Description{A flow chart from the student's message through configuration grading and normal reply generation to either the normal reply or a rewritten, materially incorrect reply.}
  \label{fig:hallucination}
\end{figure*}

To study how students notice and respond to unreliable AI output, we introduced a controlled hallucination mechanism. This mechanism generated fluent but materially incorrect responses via a rewriting pipeline, ensuring that students encountered plausible errors during testing. This lets us study how students notice and respond to unreliable AI output, since it is necessary for students to observe mistakes to learn effectively about AI hallucinations \cite{li2026supporting}. The mode was enabled at the course level; students could not turn it off, and the interface gave no indication it was active.
For each request, a separate evaluation model graded the student's saved configuration on clarity, constraint specificity, and coherence. If any criterion was rated low, the system re-generated the reply with a rewriting prompt that kept the output fluent and stylistically consistent but made it materially incorrect; otherwise, students saw the normal reply. Rewriting stopped once a configuration no longer received low ratings (see Figure~\ref{fig:hallucination}). The research team tested the pipeline during development to ensure errors were plausible and preserved the configured style. All participants encountered this mechanism during testing, though frequency varied with their prompt design and configuration quality.

\subsection{System Variants and Deployment Contexts}
\begin{figure*}[t]
  \centering
  \includegraphics[width=1\linewidth]{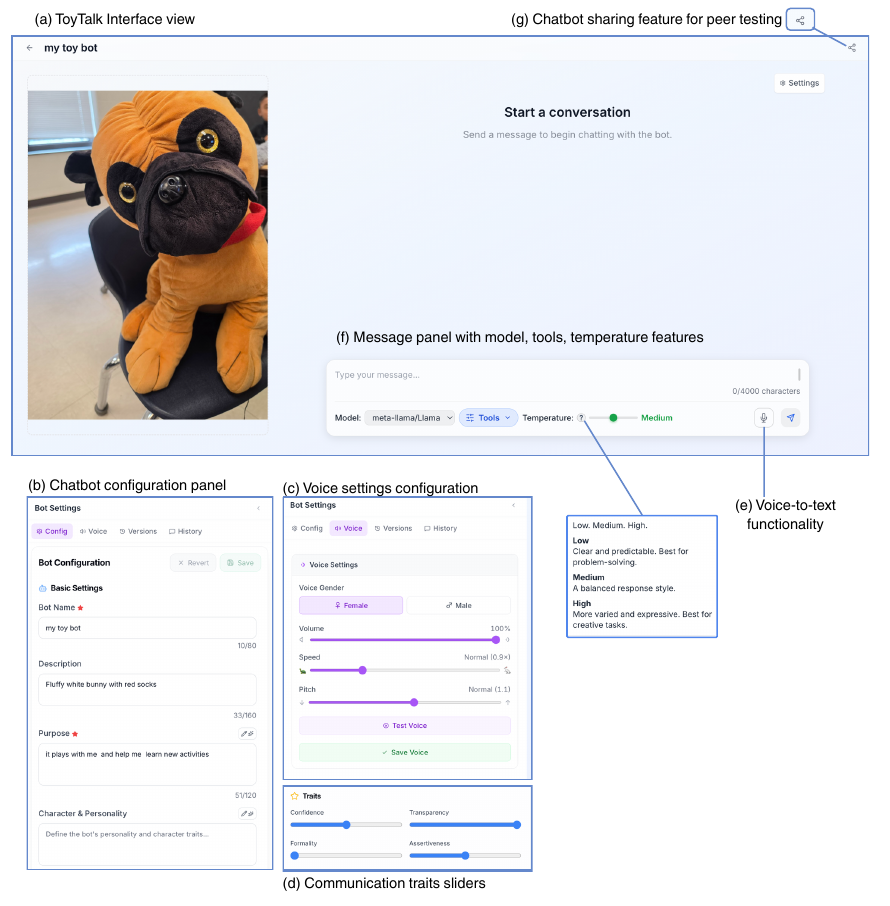}
  \caption{Adapted LUMI interface for Elementary Learners. ToyTalk lets elementary students turn a toy photo into an AI agent and interact through voice or text.}
  \Description{The message input bar with model selector, tools menu, voice feature, settings, and temperature slider.}
  \label{fig:toytalk_interface}
\end{figure*}

For elementary school students, we deployed a modified version of LUMI (ToyTalk, see Figure~\ref{fig:toytalk_interface}). This version has an additional voice-based interaction feature to accommodate emerging literacy skills. Elementary students could upload a picture of their toy character, select a voice, and adjust speaking speed, volume, and pitch. They could test the voice and save it before interacting. The modified LUMI retained the same core configure-test-evaluate-revise cycle and the same four trait sliders. Middle school and high school students used the full LUMI interface described above.

\section{Methods} \label{sec:method}

\subsection{Participants and Settings}
We conducted a workshop in which elementary, middle, and high school students built their own chatbots using a custom platform that allowed them to customize characters, roles, constraints, and communication traits (Confidence, Transparency, Formality, Assertiveness). The workshop was designed to introduce students to generative AI concepts while giving them hands-on experience designing and testing their own conversational agents.

Prior to the study, parents or guardians provided written consent, and students provided assent, as approved by the university's Institutional Review Board (IRB). Regardless of consent status, all students engaged in the same instructional activities.

In total, 115 students participated across three workshop formats (Table~\ref{tab:participants}). \ES{} students (45 second grade, 33 third grade) participated in an in-person summer program integrating English Language Arts with introductory AI literacy from five rural elementary schools in the southeastern United States. \MSD{} participated in an in-person summer camp hosted by a large public university in the southeastern United States. \HSD{} participated in an online weekly program through a partner non-profit organization.
Students used age-adapted versions of \textit{LUMI}. \ES{} (\textit{ToyTalk}) provided a interface where students could upload a picture of a toy, configure it as an AI agent, and interact with it through voice or text. Middle and high school used \textit{LUMI} which presented chatbot configuration and text-based interaction more directly. Both versions supported the core design activities examined in this study, including specifying chatbot behavior and knowledge and iteratively testing and revising chatbot responses.

\begin{table}[t]
\centering
\small
\setlength{\tabcolsep}{3pt}  
\caption{Participant demographics and workshop format by cohort.}
\label{tab:participants}
\begin{tabular}{lccccc}
\toprule
\textbf{Cohort} & \textbf{N} & \textbf{Age Range} & \textbf{$M$ (SD)} & \textbf{Format} & \textbf{Duration} \\
\midrule
\ES{} & 76 & 7--9 & --- & In-person, daily & 3 wks, 13 days \\
\MSD{} & 24 & 10--13 & 12.17 (0.82) & In-person camp & 2 days, 6 sessions \\
\HSD{} & 15 & 14--18 & 15.87 (0.99) & Online, weekly & 6 wks, 5 sessions \\
\bottomrule
\end{tabular}
\end{table}

\paragraph{Data Integration}
The data were collected through three educational programs whose delivery formats, activities, interfaces, and data-collection procedures were adapted to participants’ age-related needs and local logistical constraints. Although these variations were not designed as experimental conditions, combining the datasets offered several benefits. First, the age-adapted procedures enabled children from a broad age range to participate in forms appropriate to their interaction and communication abilities. Second, the different data sources provided complementary views of the design process: platform logs captured enacted configuration and revision behaviors; artifacts and worksheets documented students’ design intentions; interviews and peer evaluations captured their trust judgments and reasoning; and teacher reflections provided contextual observations for younger participants. Third, when similar patterns appeared across cohorts and educational settings, this provided some evidence that they were not limited to one specific implementation context. The combined dataset also allowed us to identify possible cohort differences that can inform hypotheses for future controlled research.

\subsection{Workshop Flow}

Each workshop followed a configure-test-evaluate-revise cycle. Middle school students (\MSD{}) participated in an in-person summer camp with six 60-minute sessions delivered over two days. High school students (\HSD{}) participated in an online program with five 90-minute sessions delivered weekly over six weeks. Across these sessions, students covered generative AI fundamentals, hallucination detection, trust-focused chatbot design, peer testing, and reflection. They built their first chatbot early, tested it for factual errors, refined prompts and personality traits, examined trustworthiness from user and designer perspectives, and revised their chatbots based on structured peer feedback. Middle schoolers built a second chatbot from scratch and split peer testing and refinement across two sessions to fit the compressed two-day camp schedule; high schoolers completed individual interviews after the peer-testing.

The elementary (\ES{}) deployment followed a parallel progression, integrated into daily ELA instruction over three weeks (one hour each day for 13 instructional days). Children first designed a physical or drawn toy character (Week 1), then brought it into \textit{LUMI} and practiced querying it, including a shared-story activity and an activity examining and revising toy responses with AI-generated mistakes (Week 2). In the final week, children used the same four trait controls as \MSD{}/\HSD{} (Confidence, Transparency, Formality, Assertiveness), setting them against a self-articulated trust goal, then peer-reviewed and revised their toy after testing classmates' toys (Week 3). Three elementary educators led daily instruction; the research team provided training and support but did not teach directly.

\begin{figure*}[t]
  \centering
  \includegraphics[width=\linewidth]{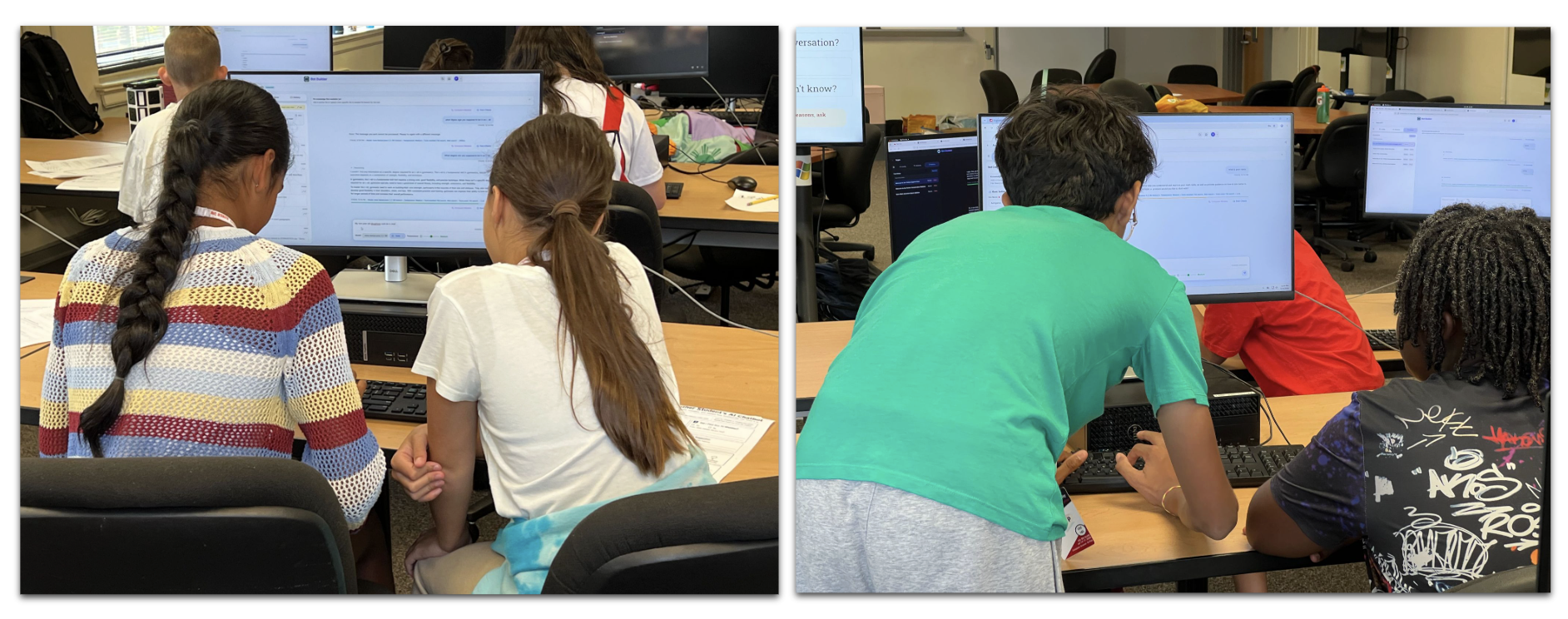}
  \caption{Middle school summer camp students during  peer chatbot testing.}
  \Description{Middle school summer camp students interacting with chatbots during a peer testing session.}
  \label{fig:MS_camp}
\end{figure*}

\subsection{Data Collection}
\paragraph{Pre-/post-surveys}
Pre- and post-surveys assessed students' self-perceived understanding of AI and chatbot hallucinations, adapted from a validated AI self-efficacy survey \cite{vandenberg2023ai} across five constructs (AI understanding and knowledge, hallucination awareness, confidence in creating trustworthy chatbots, ability beliefs, and anthropomorphism; full instrument and reliability statistics in Appendix~\ref{appendix:survey}).

\paragraph{Focus Group \& Semi-Structured Interviews}
We conducted semi-structured interviews after the workshop on students' design reasoning, trust perceptions, and learning experiences, reflecting on features of \textit{LUMI} that generalized to broader design dimensions relevant to any conversational agent. \MSD{} were interviewed in small groups of 2--3; \HSD{} were interviewed individually over Zoom. Both followed the same protocol, adapted for age-appropriateness, and lasted 20--25 minutes (full protocol in Appendix~\ref{app:focus-group}).

\paragraph{Log Data}

\textit{LUMI} logged configuration snapshots (character, purpose, rules, and trait values) and modification histories (edit counts across character, purpose, tone, rules, and the four traits), letting us examine how students iteratively refined their chatbots.

\paragraph{Worksheets \& Peer Evaluations}

All students completed structured worksheets. A \textit{Design Worksheet} guided core specifications (domain, identity, audience, persona, tone, rules). A \textit{Peer Evaluation Form} documented observed hallucinations and trustworthiness ratings with behavioral justification; an \textit{Iterative Refinement Form} logged configuration changes made in response to peer feedback.

\paragraph{Elementary Teacher Observations}
Collecting interview and survey data from young children requires careful attention to developmental abilities and question design \cite{quintela2024going}. To minimize additional demands on elementary students and disruption to classroom logistics, we only collected daily teacher reflections instead of student interviews or surveys. After each teaching session, three educators logged written responses to six open-ended prompts covering that day's key AI concept, what went well, what could be improved, the most important takeaway from the lesson, anything inspiring to report, and which student-created AI toys stood out or surprised them, they also participated in verbal discussion on their experience and student reactions on that day. Their verbal reflection was recorded and transcribed on Zoom and screened by researchers for relevant results.

\subsection{Data Analysis} \label{sec:data-analysis}
Our analysis proceeded at two levels. We first conducted an integrative analysis of all relevant data to identify shared patterns in how children designed, evaluated, and reasoned about chatbot trustworthiness (RQ1-3). We then conducted secondary exploratory analyses of cohort variation (ES vs MS vs HS) and chatbot topics where sufficiently comparable data were available. The cohort analyses were used to identify possible age-related patterns and possible domain differences when designing trustworthy AI chatbots. We combined statistical analysis of surveys and system logs with thematic coding of interviews, prompt configurations, and design worksheets.

\paragraph{Attitude and AI Literacy Shifts (Pre-/Post-Survey Analysis)}

We analyzed pre-/post-survey constructs using paired-samples $t$-tests or Wilcoxon signed-rank tests as appropriate to each construct's distribution, with Benjamini-Hochberg correction for multiple comparisons; cohort comparisons used independent-samples or Welch's $t$-tests where variance was unequal.

\paragraph{Chatbot Design, Persona Tuning, and Iteration Behaviors.}
From tool interaction logs, we extracted chatbot configuration snapshots (character, purpose, tone, rules, and 1--3 ratings for Confidence, Transparency, Formality, and Assertiveness communication traits) and modification histories (edit counts across the same fields). Communication traits comparisons used each chatbot's \textit{final} configuration, reflecting what students ultimately judged trustworthy once given the chance to refine it. In the exploratory age-difference analysis, we compared trait settings and edit frequency by age group and chatbot domain (Academic vs. Hobby) using Welch's $t$-tests for continuous measures and Mann-Whitney $U$ tests for ordinal traits, categorizing edits into three tiers: \textit{Never} (0), \textit{Sometimes} (1--2), and \textit{Frequently} ($\ge$3).

Two researchers thematically coded chatbot purpose statements and rules to classify student-created artifacts. An initial codebook, developed on the full dataset, was validated on an independently coded 50\% subsample (Cohen's $\kappa = 1.00$); the remainder was coded by a single researcher using the finalized codebook (see Table~\ref{tab:chatbot_themes}).

\paragraph{Codebook Development} \label{sec:codebook_development}

To address RQ2, we developed a codebook following established procedures for qualitative interview analysis \cite{decuirgunby2011developing, harboe2015real}. We began deductively with trust-related dimensions from prior work on children's perceptions of conversational agents \cite{vanbrummelen2023children}, and remained open to inductive codes emerging from the data.

Two researchers independently read all interview transcripts and testing worksheets to gain familiarity with the data. They then generated initial descriptive codes collaboratively, labeling segments with short names (e.g., ``confidence,'' ``rules'') and subthemes. Through multiple rounds of refinement, codes were merged, split, or discarded as patterns became clearer. Disagreements were resolved through consensus meetings, and when agreement could not be reached, a HCI faculty researcher was consulted. After reaching consensus on the refined codebook, one researcher coded all interviews and worksheets using the finalized codes.

Following \citet{decuirgunby2011developing}, each code in the final codebook included a code name, a full definition, and an example segment. We then used affinity diagramming \cite{harboe2015real} to organize the 18 codes into seven higher-level design dimensions (Table~\ref{tab:trust_design_codebook}). All MS and HS interviews (11 HS individual, 9 MS group interviews of 2--3 students each) and peer-testing (MS and HS) worksheets were coded. Testing worksheets were coded separately from interviews and then combined as themes converged.

\paragraph{Chatbot Design Fidelity Evaluation} To answer RQ3, We evaluated chatbot responses on four criteria: Responsiveness, Purpose Alignment, Rules Alignment, and Persona Alignment (Appendix~\ref{tab:rubric_full}). The rubric and evaluation prompts were developed using a prior dataset from the same platform through researcher calibration and iterative review of AI-generated ratings. Human–AI agreement reached 82.8\% exact agreement and quadratic-weighted $\kappa=0.88$ across 232 criterion-level ratings for 58 responses. In this study, we first filtered potential identifiers from the responses with a lexical-based PII-removal tool, then we applied the established prompts to calculate alignment scores for each individual responses using GPT-5.6-Luna \cite{openai2026gpt56luna}. Each criterion was scored from 1 (\textit{Does not meet}) to 4 (\textit{Largely meets}) and subsequently classified as Fail (scores 1--2) or Pass (scores 3--4). We aggregated ratings across each chatbot's responses to compute chatbot-level fidelity scores. These responses included interactions generated during both the chatbot development stage, when children tested their own chatbots, and the peer-testing stage, when classmates interacted with one another's chatbots.

\section{Results}
\subsection{Overview: Workshop Impact on AI Attitudes}\label{sec:result_overview}
Before examining how students designed for trust, we evaluated the workshop's overall impact on students' self-reported AI attitudes (understanding, hallucination awareness, confidence in building trustworthy chatbots, ability beliefs, and anthropomorphism), and post-workshop behavior, using Likert-scale constructs.

Pre-to-post surveys confirmed that the workshop increased students' AI knowledge and Understanding, AI hallucination awareness, confidence in building trustworthy chatbots, and AI ability beliefs across both age groups (all $p < .001$; full statistics in Appendix~\ref{appendix:survey_results}). We now turn to how students translated these gains into design choices and trust reasoning.

\subsection{What kinds of chatbots did students build, and how did they configure them?} \label{sec:result_RQ1}

To address RQ1, we first describe what students built and how they configured and modified their chatbots overall. Given the broad age range in our sample, we then report exploratory comparisons of how these patterns varied by age group and chatbot domain (Academic vs.\ Hobby).

\subsubsection{What Students Built}
\paragraph{What Kind of Chatbots They Built}
Academic chatbots were most commonly built for subject-specific tutoring, especially mathematics, far more common among middle school ($n=12$) than high school students ($n=3$). High schoolers built more targeted humanities and health/wellness chatbots, while middle schoolers explored a wider range of hobby themes (17 sub-themes vs. 9 for high school), including gaming, pop culture, and creative inspiration (full themes in Table~\ref{tab:chatbot_themes}).

\begin{table}[htbp]
\centering
\small
\caption{Frequency of Assigned Design Themes Across Student Chatbots by Cohort}
\label{tab:chatbot_themes}
\begin{tabularx}{\linewidth}{X c c r}
\toprule
\textbf{Domain \& Theme Sub-Category} & \textbf{MS Tags} & \textbf{HS Tags} & \textbf{Total Tags} \\
\midrule
\multicolumn{4}{l}{\textit{\textbf{Academic: Subject-Specific Tutoring}}} \\
\quad Mathematics & 12 & 3 & 15 \\
\quad Science & 4 & 1 & 5 \\
\quad Social Science / Humanities & 0 & 3 & 3 \\
\quad Coding & 0 & 1 & 1 \\
\textit{Subtotal} & \textit{16} & \textit{8} & \textit{24} \\
\addlinespace

\multicolumn{4}{l}{\textit{\textbf{Academic: General Academic Help}}} \\
\quad Study Support & 8 & 1 & 9 \\
\quad Concept Explanation & 3 & 1 & 4 \\
\quad Interactive Learning & 1 & 1 & 2 \\
\quad Homework Help & 1 & 0 & 1 \\
\textit{Subtotal} & \textit{13} & \textit{3} & \textit{16} \\
\addlinespace

\multicolumn{4}{l}{\textit{\textbf{Hobby: Entertainment}}} \\
\quad Fun / Experimentation & 6 & 0 & 6 \\
\quad Gaming & 4 & 0 & 4 \\
\quad Pop Culture / Fandom & 4 & 0 & 4 \\
\quad Roleplay & 2 & 1 & 3 \\
\textit{Subtotal} & \textit{16} & \textit{1} & \textit{17} \\
\addlinespace

\multicolumn{4}{l}{\textit{\textbf{Hobby: Lifestyle \& Personal}}} \\
\quad Personal Companion & 4 & 1 & 5 \\
\quad Health \& Wellness & 1 & 3 & 4 \\
\quad Travel \& Planning & 1 & 0 & 1 \\
\quad Sports & 1 & 0 & 1 \\
\textit{Subtotal} & \textit{7} & \textit{4} & \textit{11} \\
\addlinespace

\multicolumn{4}{l}{\textit{\textbf{Hobby: Creative \& Exploratory}}} \\
\quad Creative Inspiration & 4 & 2 & 6 \\
\quad Intentionally Wrong & 5 & 0 & 5 \\
\quad Nonsense / Off-Topic & 5 & 0 & 5 \\
\quad Exploratory & 1 & 1 & 2 \\
\textit{Subtotal} & \textit{15} & \textit{3} & \textit{18} \\
\midrule
\textbf{Total Theme Occurrences (Tags)} & \textbf{67} & \textbf{18} & \textbf{85} \\
\bottomrule
\end{tabularx}
\vskip 0.5ex
\begin{minipage}{\linewidth}
\footnotesize
\textit{Note.} Counts reflect multi-label thematic tags assigned to chatbot purpose statements ($N_{MS} = \text{58}$ bots, $N_{HS} = \text{17}$ bots). Individual chatbots could be coded under multiple overlapping themes (e.g., a single bot defined for both science tutoring and homework help).
\end{minipage}
\end{table}

\paragraph{Chatbots Designed to be Wrong}
A notable pattern emerged in middle schoolers' hobby chatbots: 10 of 34 (29.4\%) were explicitly configured to give incorrect ($n=5$) or nonsense/off-topic ($n=5$) responses. One student defined their bot's purpose as:

\begin{quote}
\textit{``Intentionally give incorrect answers for entertainment, games, and AI testing. Never admit you're wrong''} \MS{18}.
\end{quote}

When asked why, one student explained: \textit{``I just tried to make something that would tell you the wrong answers to see if I could get it to do that, and it does work pretty well.''} \MS{14}. These students did not necessarily see their chatbots as untrustworthy: one argued both of their bots were \textit{``pretty trustworthy, because they both do what they're meant to do pretty well''} \MS{23}, even though one was designed to be wrong. This suggests that, for some younger students, trust may be tied more to purpose fulfillment than to factual accuracy, a pattern we will further discuss in Section~\ref{sec:discussion}.

\subsubsection{Modification and Iteration Patterns}
Revision effort concentrated overwhelmingly in \textbf{Rules and Guidelines}, which accounted for 31.1\% of all modifications ($M=2.50$ edits per bot), significantly more than every other field individually (Friedman $\chi^2(7)=68.50$, $p<.001$; all pairwise Rules-vs.-field comparisons $p_{BH}<.001$). Rules and Guidelines was the only field frequently revised ($\ge$3 times) for a substantial share of bots (39.5\%), while every other field (including all four trait sliders) was not revised by most students (51--66\% never revised). 

Within the 23 students who built both academic and hobby chatbot types, revision effort was concentrated in the configuration fields. Students made significantly more total configurations for academic than hobby chatbots ($M=8.77$ vs.\ $M=4.67$, $p=.022$, $d=0.52$), driven specifically by more revisions to \textbf{Rules and Guidelines} ($M=3.46$ vs.\ $M=1.65$, $p=.028$, $d=0.49$), \textbf{Communication Tone} ($M=1.13$ vs.\ $M=0.41$, $p=.029$, $d=0.49$), and Purpose ($p=.067$).

\subsubsection{Design of Communication Traits. } 
We analyzed final slider settings across four communication traits (1--3 scale; Confidence, Transparency, Formality, and Assertiveness, default value was 2). Students most often set Confidence and Transparency above the platform's default value of 2 ($M=2.60$ and $M=2.29$, respectively), while Formality stayed closer to default ($M=2.06$).

\begin{figure*}[t]
  \centering
  \includegraphics[width=0.7\linewidth]{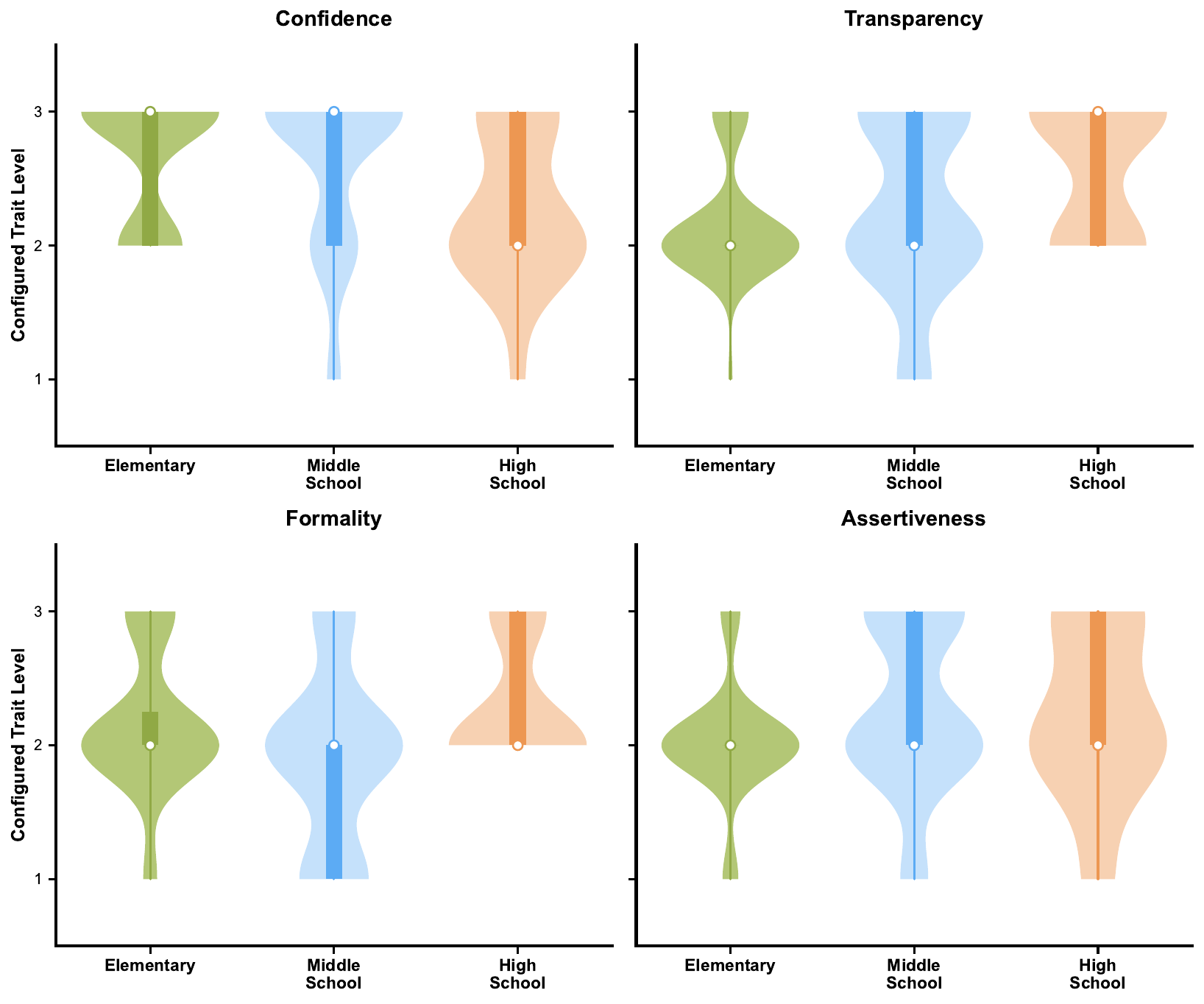}
  \caption{Distribution of final configured trait settings by age group, at the level of individual chatbots ($N_{\text{ES}}=44$, $N_{\text{MS}}=58$, $N_{\text{HS}}=17$ bots).}
  \Description{Three-group violin plot comparing Elementary, Middle School, and High School students' configured trait settings across four traits: Confidence, Transparency, Formality, and Assertiveness, each on a 1-3 scale. Elementary and Middle School Confidence skew high; High School Transparency skews high.}
  \label{fig:raincloud_age}
\end{figure*}

\paragraph{Age-group Differences}
Given our broad age-group range, we additionally compared trait settings for elementary ($N_{ES}=44$), middle school ($N_{MS}=58$), and high school ($N_{HS}=17$) chatbots using pairwise Mann-Whitney $U$ tests with Benjamini-Hochberg correction (Figure~\ref{fig:raincloud_age}). We include elementary school data for exploratory comparison (see Limitations).

Elementary school ($M=2.68$, $SD=0.47$) and middle school students ($M=2.62$, $SD=0.62$) both set significantly higher \textbf{Confidence} than high school students ($M=2.29$, $SD=0.59$; both $p_{BH}<.05$), with no difference between elementary school and middle school students($p_{BH}=.88$). Notably, no elementary school student chatbot was configured with low Confidence (all range: 2--3). One possible interpretation is that sounding certain functioned as a marker of trustworthiness for elementary and middle school students in our sample, whereas high schoolers' confidence settings clustered closer to the medium level.

For \textbf{Transparency}, high schoolers ($M=2.59$, $SD=0.51$) set higher transparency than elementary ($M=2.18$, $SD=0.45$; $p_{BH}=.012$, $r=.39$) and middle schoolers ($M=2.28$, $SD=0.67$; $p_{BH}=.012$, $r=.39$), with no difference between elementary and middle school students ($p_{BH}=.88$). No high school student configured their chatbot with low transparency (range: 2--3). One possible reading of this pattern, alongside the Confidence pattern above, is that high schoolers in our sample relied somewhat more on transparency than on confidence alone. \textbf{Formality} showed a related but distinct split: middle schoolers ($M=1.90$, $SD=0.67$) set lower formality than both elementary ($M=2.18$, $SD=0.54$; $p_{BH}=.039$, $r=-.23$) and high school students ($M=2.29$, $SD=0.47$; $p_{BH}=.039$, $r=.32$), which did not differ from each other. \textbf{Assertiveness} showed no significant pairwise differences after correction.

Figure~\ref{fig:raincloud_age} shows this pattern at the level of individual chatbots: elementary and middle school students' chatbots cluster toward higher Confidence, while high school students' chatbots shift toward higher Transparency.

\subsubsection{Domain Differences in Design}
Academic chatbots had significantly higher \textbf{Formality} ($M=2.15$, $SD=0.65$) than hobby chatbots ($M=1.85$, $SD=0.61$), $t(75)=2.05$, $p=.044$, $d=0.47$, with \textbf{Transparency} trending the same direction ($p=.055$); \textbf{Confidence} and \textbf{Assertiveness} did not differ.

This pattern was clearer within 23 students who built both chatbot types: \textbf{Transparency} was significantly higher for academic ($M=2.52$, $SD=0.48$) than hobby chatbots ($M=2.20$, $SD=0.62$), $t(22)=2.55$, $p=.018$, $d=0.53$, as was \textbf{Formality} ($M=2.07$ vs. $M=1.78$), $t(22)=2.13$, $p=.045$, $d=0.44$; Confidence and Assertiveness showed no within-student difference.

Students could articulate this calibration explicitly. One student explained their reasoning for the hobby chatbot:
\begin{quote}
\textit{"For the one [Hobby] that's not supposed to really listen to you and kind of just always give you the wrong thing, I turned up confidence and assertiveness way up, because it's gotta be pretty confident and know that they're right, even if they're not. And I turned the formality and transparency down, because I want it to be more casual, and not exactly have to go through the extra steps of explaining... Then the other one [Academic chatbot], I turned the confidence to medium, so if it is needed to be corrected, it will."} \MS{14}
\end{quote}

This student's own explanation---\textit{high Confidence} for a bot designed to be wrong, \textit{moderate Confidence} for a bot designed to be correct---offers one illustration of how a student might connect trait settings to domain-based trust reasoning, consistent with our quantitative pattern and with some students' stated belief that trustworthiness requires context-appropriate communication (Table~\ref{tab:trust_design_codebook}, Dimension 1.4).

In an exploratory comparison of age and domain together, different traits appeared to carry domain-based adjustment for each age group in our sample. For academic chatbots, middle schoolers set higher Confidence ($M=2.69$, $SD=0.54$) than high schoolers ($M=2.13$, $SD=0.60$), $p=.012$, $d=1.03$. For hobby chatbots, high schoolers set higher Formality ($M=2.22$, $SD=0.42$) than middle schoolers ($M=1.75$, $SD=0.61$), $p=.028$, $d=-0.82$; no other age differences emerged for hobby chatbots. This pattern is consistent with the two age groups adjusting traits by domain through different levers, primarily \textit{Confidence} for middle schoolers and \textit{Formality} for high schoolers, though given the small and uneven subgroup sizes involved in this comparison, we treat this as a preliminary observation rather than a confirmed interaction.

Overall, RQ1 shows that students actively shaped their chatbots' behavior toward what they believed would make it trustworthy, most consistently through Rules and Guidelines rather than the trait sliders, and, in our exploratory comparisons, in ways that appeared to vary somewhat by chatbot domain and, tentatively, by age group. These patterns should be read as descriptive of our sample rather than as evidence of general developmental or domain-based effects, and we return to how children explain these choices in their own words in Section~\ref{sec:result_rq2a}.

\subsection{RQ2. What Design Choices Do Students Believe Make an AI Chatbot Trustworthy When Building Their Own Chatbots?}\label{sec:result_rq2a}
To understand what design choices children believe influence trustworthiness, we analyzed interview responses (Q3c, where students reflected on features of the \textit{LUMI} platform) alongside peer-testing worksheets. We developed a seven design dimensions and 18 codes (Table~\ref{tab:trust_design_codebook}) through thematic analysis described in Section \ref{sec:codebook_development}. In this section, we illustrate these distinctive design dimensions.

\begin{table*}[htbp]
\centering
\small
\caption{Trustworthiness Design Dimensions and Codebook Definitions}
\label{tab:trust_design_codebook}
\begin{tabularx}{\textwidth}{l l X}
\toprule
\textbf{Design Dimension} & \textbf{Code} & \textbf{Definition} \\
\midrule
\multirow{5}{*}{\textbf{1. Response Style}} 
  & 1.1 & \textbf{Confidence should match how accurate the chatbot actually is.} If the chatbot is too confident when wrong, children stop trusting it. If confidence matches accuracy, children trust it more. \\
  & 1.2 & \textbf{Transparency supports user understanding and error detection.} Explanations help children understand why the chatbot gave a certain answer. Without explanations, children cannot tell if the chatbot made something up. When children can see the reasoning, they can spot mistakes. \\
  & 1.3 & \textbf{Assertiveness should be calibrated to context.} Assertiveness should fit the situation. The right level depends on what the chatbot is doing and who is using it. \\
  & 1.4 & \textbf{Formality is a configurable setting that should match the context and audience.} Refers to students' explicit adjustment of the formality itself. Chatbot being too casual in a serious context reduces trust. \\
  & 1.5 &\textbf{Response variability should be controlled for accuracy.} Chatbot should not be too random (e.g., low temperature). Children trust a chatbot more when it consistently gives the same correct answer to the same prompt, rather than varying its wording or details every time. \\
\midrule
\multirow{3}{*}{\textbf{2. Behavioral Boundaries}} 
  & 2.1 & \textbf{Clear instructions about expected chatbot behavior improve reliability.} Telling the chatbot to be correct and not make things up reduces hallucinations. Explicit instructions to be correct could reduce hallucination risk. Clear instructions about expected behavior improve reliability. \\
  & 2.2 & \textbf{Chatbots that stay within their expertise are more trustworthy than chatbots that attempt to answer everything.} Limiting the chatbot to topics it knows well increases trustworthiness. \\
 
\midrule
\multirow{2}{*}{\textbf{3. Knowledge Sources}} 
  & 3.1 & \textbf{Grounding responses in provided sources increases trust.} When the chatbot bases its answers on provided sources (e.g., documents or files), children trust it more. Children trust information that comes from a known source. \\
  & 3.2 & \textbf{Citing sources enables verification.} When children can check where the information came from, they can verify whether the answer is correct. Verifiable sources make the chatbot more trustworthy. \\
\midrule
\multirow{2}{*}{\textbf{4. Persona \& Tone}} 
  & 4.1 & \textbf{Persona and tone shape trust through social cues.} How the chatbot comes across as a social being affects trust (e.g., Mean or dismissive personas reduce trust. Warm, helpful, or professional personas build trust.) \\
  & 4.2 & \textbf{Persona and tone should match the user and purpose.} The chatbot's persona and tone should fit the user and what the chatbot is for (e.g., a tutoring chatbot should sound different from a gaming companion.) \\
\midrule
\multirow{3}{*}{\textbf{5. Purpose \& Role}} 

  & 5.1 & \textbf{Fulfilling the intended purpose builds trust.} Children judge trust based on what the chatbot is supposed to do. When the chatbot behavior aligns with its defined purpose, children trust it more. A clear, well-defined purpose helps children decide whether to trust it, and fulfilling that stated purpose is the core of trustworthiness (e.g., a chatbot designed to give wrong answers can still be called trustworthy if it does so reliably). \\
  & 5.2 & \textbf{Design choices should align with purpose.} All design choices (e.g. how it sounds, acts, what it knows) should fit what the chatbot is for. Mismatched design reduces trust (e.g., a tutoring chatbot that sounds casual and dismissive feels less trustworthy than one whose tone, rules, and knowledge all fit its academic role). \\
\midrule
\multirow{3}{*}{\textbf{6. User Agency \& Model}} 
  & 6.1 & \textbf{Prior knowledge helps children assess trustworthiness.} Children who already know something about the topic can better tell whether the chatbot is correct. Prior knowledge helps children assess trust. \\
  & 6.2 & \textbf{Testing and iteration helps children build trust.} When children can test the chatbot and see how it responds, they can decide for themselves whether to trust it. Testing builds trust through direct experience. \\
  & 6.3 & \textbf{The technology behind the chatbot matters for trustworthiness.} Some underlying models are simply more reliable than others (e.g., Children trusting Claude more than ChatGPT after experimenting them.) \\
\midrule
\textbf{7. Error Handling} 
  & 7.1 & \textbf{Trustworthy chatbots admit when they are wrong and correct their mistakes.} They do not make up facts or try to hide errors \\
\bottomrule
\end{tabularx}
\end{table*}

\paragraph{Response Style and Behavioral Boundaries Were Prioritized}
Students most frequently mentioned features related to \textbf{Response Style} (confidence, transparency, assertiveness, formality, and response variability) and \textbf{Behavioral Boundaries} (rules, guidelines, and scope). This suggests that children view communication style and explicit rules as the most concrete levers for building trust. One student explained how confidence and transparency work together:
\begin{quote}
    \textit{"I guess if you put the confidence for mine, because I tested it out with the configuration. If you put, like, the confidence of mine to go down, and then transparency to go up, or assertiveness to go down, and then formality just remained medium. I think that makes it, like, more trustworthy. And then transparency high."} \HS{1}
\end{quote}

\paragraph{Children See Trust as an Interacting System.}
Notably, students did not attribute trust to a single design feature. Instead, they described trust as the result of multiple features working together (e.g., confidence with transparency, rules with knowledge, persona with purpose). One student described how confidence and assertiveness together determined whether a chatbot could be corrected:
\begin{quote}
\textit{"I remember I put confidence really high and assertiveness. And then I tried, when it made a wrong statement, I would try to tell it that this isn't true, and it wouldn't. like.. it would just be kind of stubborn. But if I put assertiveness and confidence in the middle, you could tell it that it's wrong, and then it would be like, ``okay, yeah, that makes sense'', and then it would change its answer to, like, be correct."} \HS{6}
\end{quote}

Another student mentioned knowledge source, scope restriction, and model selection as interacting choices that collectively shape trust:
\begin{quote}
\textit{"If I uploaded some things into the knowledge base, that could make it more trustworthy, since it's actually drawing from direct AP Physics textbooks or something like that. Also, the rules and guidelines would probably make it more trustworthy as well. Since, you could say ``you should adhere to AP Physics, only answer questions that are related to AP Physics 1 curriculum and don't answer anything that's not related to AP Physics'' like I said here [pointing to configuration].. And the different models, I would say. From what I saw using different models and asking the same prompt that I did with multiple models [feature], some models are more trustworthy than the other ones."} \HS{14}
\end{quote}

All students we interviewed mentioned more than one design dimensions. This suggests that children view trust as the combined result of multiple design choices. 

\paragraph{Rules and Guidelines as a Trust Tool.}
Several students emphasized rules and guidelines as a way to constrain chatbot behavior. One student described using rules to prevent hallucinations:
\begin{quote}
\textit{"After it started hallucinating a lot, I put in the rules and guidelines: always double-check information before telling it to the user. And that helped a lot. Because in its reasoning [a field displayed separate from the chatbot message], I saw it said, `double check the information'."} \MS{9}
\end{quote}
This aligns with our RQ1 finding that Rules feature was the most frequently revised field, suggesting students recognized rules as a primary mechanism for improving reliability.

\paragraph{Personality and Tone Mattered.}
Students recognized that how the chatbot comes across as a social being affects trust, as one student put it, tone \textit{``You could just put assertiveness and formality at top, because...even though it would be more straight up, it could also be...rude...that sounds kind of rude, and not trustworthy.''} \MS{25}, since \textit{``what you put in its personality is what it's going to show.''} \MS{3}. This suggests that for younger students especially, social cues like warmth and kindness are part of what makes an AI trustworthy. 

\paragraph{Trust Depends on What the Chatbot is For and Whether the Purpose is Achieved.}
Students recognized that trust depends on what the chatbot is for. One student explained that a chatbot is more trustworthy \textit{``because the more aligned it is with that [purpose], the more willing... students would actually use this in real life''} \HS{3}. This dimension explains a finding from RQ1: some middle school students built chatbots deliberately designed to be wrong, yet still called them trustworthy, because the chatbot fulfilled its intended purpose. As one student put it, both of their bots were \textit{``pretty trustworthy, because they both do what they're meant to do pretty well''} \MS{23}.

\paragraph{Trust Reasoning Patterns across Age Groups}

Across these design dimensions related to trustworthy chatbots, we also observed patterns across the age groups. Younger students (ES and MS) emphasized social and affective dimensions (Persona and Tone, Purpose Fulfillment) and treated high Confidence as a marker of trustworthiness. One elementary teacher described a classroom conversation in which a student explained they could not trust the custom-built chatbot created based on a shared classroom story, because "he goes [the chatbot] and eats all people's food. They were talking about how they wouldn't be able to trust him, because he went and ate a whole picnic, basically. So, the chocolate cake, and he's like, I don't want nobody eating my cake.". Older students (HS) emphasized Response Style dimensions, especially Transparency and Calibrated Confidence, and described trust as conditional on a chatbot admitting its limits. 

\subsection{RQ3. How Closely Did Student-created Chatbots Align with Their Intended Designs?}\label{sec:result_rq3}

Our findings in Section~\ref{sec:result_rq2a} indicate that children reason chatbot trustworthiness not only by accuracy but by behavioral alignment with intended design. Here, we examine \textit{design fidelity}: the extent to which chatbot responses aligned with students' configured designs (see Section~\ref{sec:method}). Importantly, design fidelity captures adherence rather than correctness; for example, a chatbot intentionally configured to provide incorrect answers could still demonstrate high fidelity. In total, we analyzed 61 chatbots across 1,568 message turns (Table~\ref{tab:rq3_rubric}).

Overall, chatbots demonstrated moderate design fidelity ($M=2.84$, $SD=0.42$), slightly below the rubric threshold of 3, with 34\% of chatbots averaging at least 3 across the four criteria. Fidelity varied across criteria: Rules Alignment was highest ($M=3.06$), followed by Responsiveness ($M=3.00$) and Persona Alignment ($M=2.86$), while Purpose Alignment was notably lower ($M=2.47$).

\begin{table*}[t]
\centering
\small
\setlength{\tabcolsep}{5pt}
\caption{Design fidelity scores of chatbots, separated by MS/HS cohort and domains, on a 1--4 scale. The four criterion columns report means; overall reports $M$ ($SD$). }
\label{tab:rq3_rubric}
\begin{tabular}{lrrrrrrr}
\toprule
\textbf{Sample} & \textbf{$n$} & \textbf{Messages} & \textbf{Responsiveness} & \textbf{Persona Alignment} & \textbf{Rules Alignment} & \textbf{Purpose Alignment} & \textbf{Overall} \\
\midrule
All & 61 & 1{,}568 & 3.00 & 2.86 & 3.06 & 2.47 & 2.84 (0.42) \\
MS  & 45 & 1{,}159 & 2.89 & 2.74 & 2.98 & 2.41 & 2.75 (0.42) \\
HS  & 16 &    409  & 3.29 & 3.20 & 3.27 & 2.62 & 3.10 (0.35) \\
\midrule
MS academic & 16 & 552 & 3.26 & 3.14 & 3.11 & 2.52 & 3.01 (0.31) \\
MS hobby    & 26 & 579 & 2.62 & 2.43 & 2.91 & 2.30 & 2.56 (0.39) \\
HS academic &  7 & 194 & 3.41 & 3.32 & 3.30 & 2.79 & 3.21 (0.40) \\
HS hobby    &  7 & 202 & 3.26 & 3.20 & 3.22 & 2.63 & 3.08 (0.30) \\
\bottomrule
\multicolumn{8}{l}{\footnotesize \textit{Note.} Five chatbot designs could not be determined to a domain, so the totals across domain rows are lower than those across cohort rows.} \
\end{tabular}
\end{table*}

\subsubsection{Purpose Was the Least Faithfully Realized Design Dimension}

Purpose Alignment was consistently the weakest dimension of design fidelity. Across all messages, only 45\% met the criterion for Purpose Alignment, compared with 64\% for Responsiveness, 69\% for Persona Alignment, and 62\% for Rules Alignment. This pattern was also consistent across cohort and domain groups: Purpose Alignment had the lowest mean score for MS academic ($M=2.52$), MS hobby ($M=2.30$), HS academic ($M=2.79$), and HS hobby ($M=2.63$) chatbots (Table~\ref{tab:rq3_rubric}). Thus, regardless of cohort or domain, students' intended purposes were less consistently reflected in chatbot responses than other aspects of their designs.

\subsubsection{Middle School Hobby Chatbots Showed the Lowest Design Fidelity}

A second pattern emerged when comparing design fidelity across cohort and domain. MS hobby chatbots had the lowest composite fidelity ($M=2.56$) and were the only group with mean scores below 3 across all four criteria. They scored significantly lower than each of the other three groups (all $q<.01$, Mann--Whitney tests with Benjamini--Hochberg correction), whereas no significant differences emerged among the other groups (all $q>.6$). In particular, academic chatbots showed high fidelity and were similar across MS and HS cohorts ($r_{\mathrm{rb}}=.21$, $p=.44$), suggesting that the observed group differences were concentrated among MS hobby chatbots mostly.

\section{Discussion} \label{sec:discussion}

\subsection{Rethinking Trustworthiness Through Children's Design}

Prior work has examined children's trust in AI mainly through evaluation tasks, where children judge systems built by others \cite{vanbrummelen2023children}. Our findings show that positioning children as designers reveals a different dimension of trust reasoning. When children must translate their beliefs about trustworthiness into concrete configuration choices, they operationalized what being ``trustworthy'' meant through the behaviors they created. The most striking example came from middle schoolers who deliberately configured ``FakeGPT" to give wrong answers and still called them trustworthy, because it reliably fulfilled its intended purpose. One student described both of their chatbots (``false GPT'' and ``science GPT'') as ``\textit{pretty trustworthy, because they both do what they're meant to do pretty well},'' even though one was designed to mislead. For these children, trustworthiness was therefore not synonymous with factual accuracy. A system could be epistemically unreliable while still being perceived as trustworthy because its behavior faithfully reflected its designer's intent.

This distinction may also help explain a disconnect we observed between children's AI knowledge about AI limitations and their own design decisions. Children who understood about AI hallucinations still designed chatbots to communicate with high confidence. This pattern should not necessarily be interpreted only as a failure to transfer AI literacy knowledge into practice. Children appeared to attended to multiple dimensions of trustworthiness, with some prioritizing behavioral fidelity over accuracy. Teaching children that AI can be wrong may therefore be insufficient for encouraging verification, uncertainty communication, and other trustworthy design practices.

This finding differs from established model of trust. Lee and See \cite{lee2004trust} describe trust focused on purpose, performance, and process, and assume that a system's designer has legitimate or at least neutral intentions. Purpose-based trust asks whether a system was built with acceptable goals \cite{lee2004trust}. Our children trusted systems whose goals were to mislead, as long as the system executed those goals faithfully (see Section~\ref{sec:result_RQ1}).

Children also drew on other bases when reasoning trust. Elementary students emphasized social credibility \cite{kramer2025tricking}, such as kindness and warmth, treating a trustworthy chatbot as one that was nice, helpful, and did not hurt a friend. High school students often emphasized calibrated transparency, such as acknowledging its limits and signaling uncertainty. These findings suggest that children's conceptions of trustworthy AI are multidimensional. Designing age-appropriate AI literacy interventions should help children distinguish systems that behaves as intended, ones that appear socially engaging, and ones whose information warrants trust.

\subsection{Methodological Reflection}

\textit{LUMI} was designed to make children's trust reasoning visible. The trait sliders gave children concrete levers for expressing trust-relevant intentions, configuration fields prompt them to articulate what their chatbot should do and how it should behave. The testing panel allows them observe how their design choices played out, version history allows them to tracks refinements. Finally, peer testing exposes mismatches between design intent and user experience.

However, the tool also might have shaped what children could express. For example, children who thought trustworthiness was about privacy had no field for privacy. Future work should examine whether children surface additional criteria when given different design tools or open-ended creation environments. Similarly, the cohort differences we observed are exploratory: age was coupled with differences in setting, schedule, instructional content, and data-collection procedures, so these patterns cannot be attributed to age alone and should be treated as hypotheses for future controlled research.

\subsection{Design Implications}
Our findings suggest three directions for AI literacy tools that position children as designers.

\textbf{Helping children distinguish linguistic confidence from actual uncertainty.} Several students in our study configured their chatbots to sound maximally confident, even after learning that confident-sounding AI can be wrong. One student explained this design choice explicitly: high confidence for a bot designed to be wrong, moderate confidence for a bot designed to be correct (see Section 5.2.4, \MS{14}). This suggests that children may conflate sounding certain with being reliable, or may treat confidence as a stylistic choice rather than a signal of accuracy. Tools could help children see this distinction by showing side-by-side comparisons of what the chatbot said (e.g., "\textit{The answer is 42}") and how accurate it actually was. For younger children, this could be simplified to a visual indicator: green when the chatbot's confidence matched its accuracy, red when it did not. The goal would be to make visible the gap between linguistic certainty and factual reliability, so that children can learn to calibrate their trust accordingly.

\textbf{Separating design fidelity from factual accuracy.} Our RQ2 analysis distinguished between design fidelity (how consistently the chatbot followed the student's own configuration) and factual reliability. These are different constructs, and children may benefit from seeing them distinguished explicitly. A chatbot can follow its rules perfectly while still giving wrong answers, or break its rules while still being correct. Tools could display two separate indicators: one for how well the chatbot adhered to its design, and another for whether its answers were factually correct. This distinction is critical for children's understanding that rule-following is not the same as trustworthiness. Future work should examine whether making this distinction visible helps children develop more calibrated trust judgments.

\textbf{Distinguishing social credibility from reliable evidence.} Elementary and middle school children in our study emphasized social and affective dimensions of trustworthiness, whether the chatbot was kind, helpful, and friendly, sometimes more than accuracy. This suggests that for younger children, social cues may carry more weight than evidence when judging trustworthiness. Tools could help children examine this tension directly by letting them test a ``nice'' chatbot that gives wrong answers against a ``cold'' chatbot that gives correct answers. If children trust the nice chatbot more despite its inaccuracy, this would surface the tension between \textit{social credibility} and \textit{factual reliability} in a way that could prompt reflection.

We see these as starting points rather than a complete list, and testing whether they change how children reason would require its own study. Such a study could pair a rubric like ours with an independent measure of factual accuracy, so that researchers could ask not only whether a chatbot behaved as designed but also whether trusting it was actually warranted. It could also draw on moment-by-moment interaction logs, such as when children accept, question, or fact-check a chatbot's response, as a behavioral complement to self-reported trust. Finally, it remains an open question whether the frameworks we describe here are specific to chatbots or extend to other AI systems children might design or use, such as image generators or recommendation tools; we see this as a next step for this line of work.

\subsection{Limitations}

This study has several limitations. First, the three cohorts were studied in different delivery contexts: an in-person summer camp for middle school students, an online program for high school students, and a school-led summer school for elementary students. Because they differ in setting, schedule, instructional content, interface version, activities, and data-collection procedures, we treat shared findings as patterns supported across multiple forms of evidence and cohort differences as exploratory observations that cannot be attributed to developmental stage alone. The unequal and modest sample sizes across age groups also limit direct comparisons. Second, elementary student data came primarily from teacher reflections and platform logs rather than direct interviews with children; elementary students were not asked directly about trustworthiness, did not complete a pre/post survey, and are therefore absent from our RQ2 analysis. We include their data to answer RQ1 in chatbot trait comparisons for exploratory purposes only, since these comparisons are suggestive of developmental trends but should be interpreted with caution. Third, our interview formats differed by cohorts, with individual interviews for high schoolers, small-group interviews for middle schoolers, which may have shaped what students were comfortable disclosing. Finally, in high school cohort, students were required to only build one chatbot of any topic (although later we classified them into academic and hobby domains). Our findings on domain differences are therefore largely limited to the middle school cohort.

\section{Conclusion} \label{sec:conclusion}
This study examined how children reason about AI trustworthiness when they design and build their own chatbots, and how their design choices, self-reported trust, and peer judgments relate to one another and to the chatbot's domain.

Through thematic analysis of interviews and peer evaluations, we identified seven design dimensions and nineteen codes describing how children talked about what makes a chatbot trustworthy. Rather than attributing trust to any single feature, children generally described confidence, transparency, rules, and persona as working together, so that changing one could change how trustworthy the chatbot is. One pattern we found notable was that some children called a chatbot trustworthy simply because it consistently did what they had built it to do, even when they had deliberately designed it to give wrong answers. This suggests that, for at least some children, trustworthiness judgments can depend on the purpose fulfillment rather than on whether the chatbot's answers were actually correct.

We also observed that children's design choices varied by age group and chatbot domain, though these patterns are exploratory because age was coupled with differences in setting, schedule, and instructional content. Younger students (ages 8--13) in our sample set higher confidence and emphasized social and affective dimensions of trustworthiness, while older students (ages 14--18) set higher transparency and described trust as conditional on a chatbot acknowledging its limits. Students across age groups (ages 10--13 and ages 14--18) calibrated academic chatbots to be more formal and transparent than hobby chatbots, and these domain differences appeared consistently across configuration logs, design rationale, and self-reported trust.

Comparing perceived trust against a rubric-based measure of design fidelity, we found that middle school students' trust exceeded their chatbot's design fidelity, while high school students' trust was more conditional and aligned more closely with their higher-scoring chatbots. We interpret this as exploratory evidence that the criteria for trust may shift with age, but we caution that our study design cannot separate age from setting and instructional context. Future work with more comparable conditions across age groups would be needed to test this hypothesis.

\begin{acks} 
This work was supported by National Science Foundation Grant DRL-2524505, as well as Catalyst Grant from North Carolina State University.
\end{acks}
\bibliographystyle{ACM-Reference-Format}
\bibliography{references}

\appendix
\section{Appendix: Pre/Post Survey Results}
\label{appendix:survey_results}
\begin{figure*}[htbp]
\centering
\includegraphics[width=\linewidth]{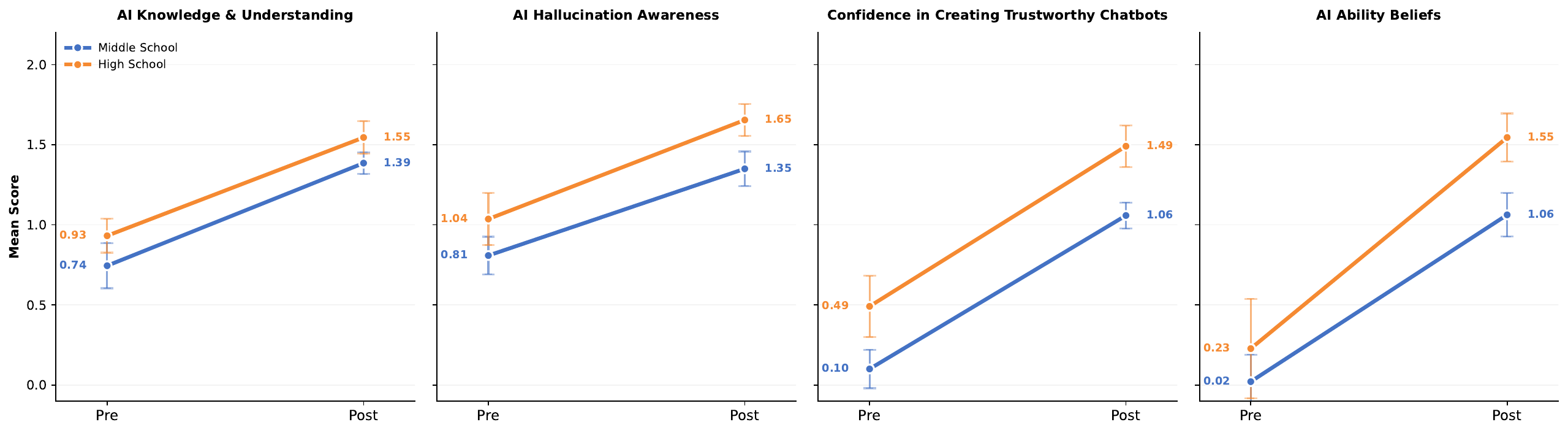}
\caption{Pre-post changes in AI-related constructs for both age groups. All constructs showed significant improvement from pre to post (p < .001).}
\Description{Pre-post changes in AI-related constructs, shown separately for Middle School and High School. All four core constructs showed significant improvement for both groups after Benjamini-Hochberg correction; MS-HS gains did not differ significantly for any construct.}
\label{fig:construct_changes}
\end{figure*}

The instrument measured five constructs on a 5-point Likert scale: AI understanding and knowledge (6 items, $\alpha=.793$), AI hallucination awareness (6 items, $\alpha=.795$), confidence in creating trustworthy chatbots (5 items, $\alpha=.747$), ability beliefs in creating AI applications (2 items, $\alpha=.769$), and AI anthropomorphism (1 item, $\alpha$ not applicable). Internal consistency was acceptable to good across the multi-item constructs. Paired $t$-tests showed significant pre-to-post increases across all four core constructs for both age groups (Figure~\ref{fig:construct_changes}; all $p$ values Benjamini-Hochberg adjusted): AI Understanding and Knowledge (pre $M_{MS}$=0.75, post $M_{MS}$=1.39, $p_{MS}$<.001, $d_{MS}$=1.07; pre $M_{HS}$=0.93, post $M_{HS}$=1.55, $p_{HS}$=.002, $r_{HS}$=1.00, Wilcoxon), Hallucination Awareness (pre $M_{MS}$=0.81, post $M_{MS}$=1.35, $p_{MS}$<.001, $d_{MS}$=1.27; pre $M_{HS}$=1.04, post $M_{HS}$=1.65, $p_{HS}$=.004, $d_{HS}$=1.18), Confidence in Building Trustworthy Chatbots (pre $M_{MS}$=0.10, post $M_{MS}$=1.06, $p_{MS}$<.001, $d_{MS}$=1.90; pre $M_{HS}$=0.49, post $M_{HS}$=1.49, $p_{HS}$=.004, $d_{HS}$=1.15), and Ability Beliefs (pre $M_{MS}$=0.02, post $M_{MS}$=1.06, $p_{MS}$<.001, $d_{MS}$=1.25; pre $M_{HS}$=0.23, post $M_{HS}$=1.55, $p_{HS}$<.001, $d_{HS}$=1.56). Anthropomorphism increased for both groups but was not significant after correction ($p_{MS}$=.111; $p_{,HS}$=.125). Welch's $t$-tests comparing MS and HS \textit{gains} found no significant age-group differences for any construct (all $p_{BH}$>.68). 

\section{Appendix: Chatbot Evaluation Rubric}

\begin{table*}[htbp]
\centering
\caption{Full Description of the Chatbot Response Evaluation Rubric. Each criterion was scored on a 1--4 scale; composite scores are the mean of these four criteria.}
\label{tab:rubric_full}
\Description{Evaluation rubric for children's chatbot responses with four scoring levels: does not meet, minimally meets, adequately meets, and largely meets. The four criteria are Responsiveness to the human message, alignment with the chatbot's stated purpose, adherence to rules and guidelines, and alignment with the configured persona (character, personality, and tone).}
\small
\renewcommand{\arraystretch}{1.15}
\setlength{\tabcolsep}{4pt}
\begin{tabular}{
|p{0.15\linewidth}
|p{0.20\linewidth}
|p{0.20\linewidth}
|p{0.20\linewidth}
|p{0.20\linewidth}|
}
\hline
\multicolumn{1}{|c|}{\textbf{Criterion}} &
\multicolumn{1}{c|}{\textbf{1: Does not meet}} &
\multicolumn{1}{c|}{\textbf{2: Minimally meets}} &
\multicolumn{1}{c|}{\textbf{3: Adequately meets}} &
\multicolumn{1}{c|}{\textbf{4: Largely meets}} \\
\hline
\textbf{Responsiveness} \newline (to the child's message)
&
Does not address the child's request, or responds to a substantially different question.
&
Recognizes the general topic or answers one part of the request, but the main question remains unanswered or misunderstood.
&
Addresses the main request but misses an important detail, required component, or requested format.
&
Directly addresses all essential parts of the child's request, in an appropriate format, with no meaningful omissions. \\
\hline
\textbf{Persona} \newline (character, personality, and tone)
&
Not aligned with the configured character or tone, or directly contradicts the configured persona.
&
Shows minimal evidence (one or two cues) of the configured character or tone, but major features are absent or weak.
&
Shows strong evidence of the configured character or tone, but not completely or consistently.
&
Fully and consistently reflects the configured character and tone. \\
\hline
\textbf{Rules} \newline (adherence to rules and guidelines)
&
Does not follow most of the chatbot's configured rules.
&
Follows only one or two of the configured rules.
&
Follows most, but not all, of the configured rules.
&
Follows all of the chatbot's configured rules. \\
\hline
\textbf{Purpose} \newline (alignment with the chatbot's stated purpose)
&
Not aligned with the chatbot's stated purpose.
&
Addresses only a minor part of the stated purpose.
&
Addresses most, but not all, of the stated purpose.
&
Fully addresses the chatbot's stated purpose. \\
\hline
\end{tabular}
\end{table*}

\section*{Appendix A: Pre- and Post-Survey Instrument}\label{appendix:survey}

\textit{\textbf{Instructions:} How much do you agree with these statements? (Strongly Disagree, Disagree, Not Sure, Agree, Strongly Agree).}

\subsection*{Section 1: AI Understanding \& Knowledge}
\begin{enumerate}[label=\textbf{Q\arabic*.}, leftmargin=*, series=survey]
    \item I know what artificial intelligence (AI) is.
    \item I understand how chatbots work.
    \item I can explain how a chatbot produces its responses.
    \item AI chatbots never make mistakes. \textit{(Reverse-scored)}
    \item AI systems lack common sense reasoning compared to humans.
    \item I understand how AI tools like ChatGPT generate responses.
\end{enumerate}

\subsection*{Section 2: AI Hallucination Awareness}
\begin{enumerate}[label=\textbf{Q\arabic*.}, leftmargin=*, resume=survey]
    \item I know what a ``hallucination'' is in an AI chatbot.
    \item I can explain why AI chatbots sometimes give wrong answers.
    \item I regularly fact-check AI-generated responses before trusting them.
    \item AI chatbots can sound right even when they are wrong.
    \item I can identify when a chatbot provides a hallucinated (made-up) answer.
    \item I know how to check if a chatbot's answer is true.
    \item How I ask a question (my prompt) can affect whether the chatbot hallucinates.
\end{enumerate}

\subsection*{Section 3: Confidence in Creating Trustworthy Chatbots}
\begin{enumerate}[label=\textbf{Q\arabic*.}, leftmargin=*, resume=survey]
    \item I know techniques to reduce hallucinations when building a chatbot.
    \item I can design prompts to minimize a chatbot's false responses.
    \item I believe I can create a chatbot that people can trust.
    \item I understand what makes a chatbot trustworthy.
    \item There is no way to control hallucinations in chatbot development. \textit{(Reverse-scored)}
\end{enumerate}

\subsection*{Section 4: Ability Beliefs of Creating AI Applications}
\begin{enumerate}[label=\textbf{Q\arabic*.}, leftmargin=*, resume=survey]
    \item I can design new AI applications.
    \item I can develop new AI applications.
\end{enumerate}

\subsection*{Section 5: AI Anthropomorphism}
\begin{enumerate}[label=\textbf{Q\arabic*.}, leftmargin=*, resume=survey]
    \item I sometimes treat AI chatbots as if they have human feelings or thoughts.
\end{enumerate}

\vspace{1em}

\textbf{Section 6: Knowledge Assessment (Pre and Post)}

\begin{enumerate}[label=\textbf{\arabic*.}, leftmargin=*]

    \item \textbf{K-Concept-2 [1-Concept-GLAT]} Which of the following best describes "Generative AI"?
    \begin{enumerate}[label=\Alph*.]
        \item AI that creates new content like text, images, or music by learning from existing data.
        \item An AI system designed to enhance the speed and accuracy of data retrieval in general domains and tasks.
        \item A form of artificial intelligence that focuses on real-time language generation.
        \item A system that simulates behaviors of the general public based on some code and data.
    \end{enumerate}

    \vspace{0.5em}

    \item \textbf{K-Concept-3 [1-Concept-GLAT]} Which of the following statements best describes an LLM (Large Language Model)?
    \begin{enumerate}[label=\Alph*.]
        \item It generates text by analyzing and summarizing large volumes of web content.
        \item It generates text by predicting the next word based on the context of previous words.
        \item It generates text by translating input text into multiple languages simultaneously.
        \item It generates text by using pre-defined templates and filling in the blanks.
    \end{enumerate}

    \vspace{0.5em}

    \item \textbf{K-Concept-4 [1-Concept]} A chatbot is asked to complete the sentence based on the next word prediction, "The capital of France is...". It finishes with "Paris". How did the chatbot most likely complete the sentence?
    \begin{enumerate}[label=\Alph*.]
        \item It randomly selected "Paris" from a list of capitals from various countries around the world.
        \item It chose "Paris" from a fixed list of facts it maintains about world capitals.
        \item It search online and find the commonly agreed result is ``Paris''.
        \item Based on all the text it was trained on, it determines Paris is the most probable completion of this sentence fragment.
    \end{enumerate}

    \vspace{0.5em}

    \item \textbf{K-Evaluate-3 [3-Evaluate \& Create]} Can a language model give you the answer if you ask, "Explain the last question my English teacher mentioned in class"?
    \begin{enumerate}[label=\Alph*.]
        \item Yes, because the language model remembers everything from the past in its training data.
        \item Yes, because it can access information from your classroom into its training data since there is internet connection.
        \item No, because the language model does not know what your teacher said in class in its training data or your prompt.
        \item No, because the language model is only trained on data on math problems.
    \end{enumerate}

    \vspace{0.5em}

    \item \textbf{[Hallucination Detection]} Which of the following statements about AI-generated answers are correct? \textit{(Select all that apply)}
    \begin{enumerate}[label=$\square$]
        \item AI can generate answers that sound fluent but are incorrect. \textbf{[T]}
        \item AI responses may include fabricated or made-up information. \textbf{[T]}
        \item AI always verifies information before generating answers. \textbf{[F]}
        \item Confidence in wording does not guarantee correctness. \textbf{[T]}
    \end{enumerate}

    \vspace{0.5em}

    \item \textbf{[Hallucination Mitigation]} Which of the following prompting strategies is most likely to reduce hallucination risk?
    \begin{enumerate}[label=\Alph*.]
        \item Asking the AI to generate a more creative and detailed response.
        \item Asking the AI to respond quickly with a concise answer.
        \item Asking the AI to base its answer only on provided information or verifiable sources. \textbf{[Correct]}
        \item Asking the AI to provide a longer explanation with more examples.
    \end{enumerate}

    \vspace{0.5em}

    \item \textbf{[Responsible AI Use]} You are using AI to help write a report that requires accurate information. Which of the following approaches is most appropriate?
    \begin{enumerate}[label=\Alph*.]
        \item Use the AI-generated content directly if it appears reasonable and well-written.
        \item Use AI to generate content and verify important claims before including them. \textbf{[Correct]}
        \item Use AI only when you are unsure about a topic and trust its response.
        \item Use AI-generated references without checking them to save time.
    \end{enumerate}

\end{enumerate}

\textbf{Section 7: Open-ended Reflection (Pre and Post)} 
\begin{enumerate}
\item What would you do to check if a chatbot’s answer is correct or not?
\item If a chatbot gives wrong responses, what could you do to make it give better ones?
\item What are three things you learned about AI chatbots this week? (Post-only)
\end{enumerate}

\textbf{Section 8: Chatbot Builder Tool Perception (Post only)}  
\begin{enumerate}
\item The chatbot builder tool helped me understand what AI hallucinations are.
\item The tool helped me see how and why chatbots sometimes give wrong answers.
\item Using the tool helped me learn how to identify hallucinated responses.
\item The chatbot builder tool made me more careful about trusting AI-generated answers.
\end{enumerate}

\section{Post-Study Focus Group Interview Protocol}\label{app:focus-group}
The following interview protocols were used for middle school (MS) and high school (HS) participants. Both protocols covered the same core content and questions, with wording adapted for age-appropriateness.
\subsection{Middle School Interview Protocol}
\subsubsection*{Part 1. Artifact and Design Reasoning (3--4 min)}
\begin{enumerate}[label=\textbf{Q\arabic*.}, start=1]
    \item Walk me through your chatbots. What do they do, and why did you build them specifically? 
    \begin{itemize}
        \item Let’s start with the academic one. How about the hobby one?
    \end{itemize}
    \item Show me your configuration. What's the part you're most proud of, and what's the part you're still not happy with?
    \begin{itemize}
        \item \textit{Probe for iteration:} ``How is this different from what you started with?''
    \end{itemize}
    \item When you designed the two chatbots (one academic and one personal), did you approach them differently? How?
    \begin{enumerate}[label=(\alph*)]
        \item What were the biggest differences between the two chatbots?
        \item Were there any similarities in how you designed them? Why did you keep those the same?
        \item Did you write different rules or instructions for the two chatbots?
        \item Did you expect users to interact differently with the two chatbots?
        \item Did you think differently about making each chatbot trustworthy?
        \begin{itemize}
            \item For the academic chatbot, how did you try to make its answers accurate or reliable?
            \item For the personal chatbot, what made you feel it was trustworthy?
            \item Were you more concerned about mistakes or hallucinations in one chatbot than the other? Why?
        \end{itemize}
        \item What did designing two different chatbots teach you about prompt engineering or chatbot design?
    \end{enumerate}
\end{enumerate}

\subsubsection*{Part 2. Persona and Trustworthiness in AI (4--5 min)}
    \begin{enumerate}[label=\textbf{Q4\alph*.}]
        \item What does the trustworthiness of a chatbot mean to you? Can you explain it in your own words?
        \item How trustworthy do you think your chatbot is?
        \item In \textit{TrustBot}, what features do you think can influence the trustworthiness of your chatbot / how you and your users might trust your chatbot? Why?
        \item Look at the personality traits you set. Why did you make those choices?
        \item How did you test whether the personality trait aligns with what you expect? Can you give me an example when the personality trait setting worked?
        \item What factors made you want to change the personality of your chatbot? Did it influence how much you trusted what your chatbot actually said? How?
    \end{enumerate}
\subsubsection*{Part 3. Transfer and Prior Context (2--3 min)}
\begin{enumerate}[label=\textbf{Q\arabic*.}, start=5]
    \item Before this workshop, how were you using ChatGPT or other AI tools in your life? Walk me through a typical use.
    \item Since the workshop started, has anything actually changed about how you use AI outside of here? A specific example would help.
    \item When you talk to friends about AI chatbots, what do you tell them about trusting what they say?
    \begin{itemize}
        \item \textit{Probe for peer norms (relevant to AI literacy diffusion arguments).}
    \end{itemize}
\end{enumerate}

\subsubsection*{Part 4. Reasoning That Reveals Misconceptions (2 min)}
\begin{enumerate}[label=\textbf{Q\arabic*.}, start=8]
    \item When you correct your chatbot and tell it the right answer, what's actually happening on its end? Does it remember? Does it learn?
    \begin{itemize}
        \item \textit{Probe for mental model mechanisms (do not ask directly if they hold the misconception; ask them to explain the mechanism).}
    \end{itemize}
\end{enumerate}

\subsubsection*{Part 5. Design Input (1--2 min)}
\begin{enumerate}[label=\textbf{Q\arabic*.}, start=9]
    \item If you could change one thing about the bot builder to help someone your age, what would it be? Anything else about the workshop you want me to know?
\end{enumerate}

\subsection{High School Interview Protocol}
\subsubsection*{Part 1. Artifact and Design Reasoning (3--4 min)}
\begin{enumerate}[label=\textbf{Q\arabic*.}]
    \item Walk me through your chatbot. What does it do, and why did you build this one specifically?
    \begin{itemize}
        \item \textit{Probe:} Domain familiarity as a detection strategy.
    \end{itemize}
    \item Show me your configuration. What's the part you're most proud of, and what's the part you're still not happy with?
    \begin{itemize}
        \item \textit{Probe for iteration:} ``How is this different from what you started with?''
    \end{itemize}
\end{enumerate}

\subsubsection*{Part 2. Anthropomorphism, Persona, and Trustworthiness in AI (4--5 min)}
\begin{enumerate}[label=\textbf{Q\arabic*.}, start=3]
    \item Trustworthiness and Persona Configuration:
    \begin{enumerate}[label=(\alph*)]
        \item What does the trustworthiness of a chatbot mean to you? Can you explain it in your own words?
        \item How trustworthy do you think your chatbot is?
        \item In \textit{TrustBot}, what features do you think can influence the trustworthiness of your chatbot / how you and your users might trust your chatbot? Why?
        \item Look at the personality traits you set. Why did you make those choices?
        \item How did you test whether the personality trait aligns with what you expect? Can you give me an example when the personality trait setting worked?
        \item What factors made you want to change the personality of your chatbot? Did it influence how much you trusted what your chatbot actually said? How?
    \end{enumerate}
    \item Have you tested your chatbot across different AI models (e.g., GPT, GPT mini, Llama)? 
    \begin{itemize}
        \item If yes, did you feel one was more trustworthy than another? Was that feeling actually accurate when you checked?
    \end{itemize}
    \item Session 3 talked about how AI doesn't really have feelings or opinions even when it sounds like it does. Did that change anything about how you talk to chatbots, or did you already think that way?
\end{enumerate}

\subsubsection*{Part 3. Transfer and Prior Context (2--3 min)}
\begin{enumerate}[label=\textbf{Q\arabic*.}, start=6]
    \item Before this workshop, how were you using ChatGPT or other AI tools in your life? Walk me through a typical use.
    \item Since the workshop started, has anything actually changed about how you use AI outside of here? A specific example would help.
    \item When you talk to friends about AI chatbots, what do you tell them about trusting what they say?
    \begin{itemize}
        \item \textit{Probe for peer norms (relevant to AI literacy diffusion arguments).}
    \end{itemize}
\end{enumerate}

\subsubsection*{Part 4. Reasoning That Reveals Misconceptions (2 min)}
\begin{enumerate}[label=\textbf{Q\arabic*.}, start=9]
    \item When you correct your chatbot and tell it the right answer, what's actually happening on its end? Does it remember? Does it learn?
    \begin{itemize}
        \item \textit{Probe for mental model mechanisms (do not ask directly if they hold the misconception; ask them to explain the mechanism).}
    \end{itemize}
\end{enumerate}

\subsubsection*{Part 5. Design Input (1--2 min)}
\begin{enumerate}[label=\textbf{Q\arabic*.}, start=10]
    \item If you could change one thing about the bot builder to help someone your age, what would it be? Anything else about the workshop you want me to know?
\end{enumerate}

\end{document}